\documentclass[usenatbib,usedcolumn]{mn2e}
\usepackage{psfig,epsf}
\usepackage{multirow}

\newcommand{\Msolar}{\mbox{${\; {\rm M_{\sun}}}$}}

\newcommand{\hi}{\mbox{H$\,${\sc i}}}

\newcommand{\heii}{\mbox{He$\,${\sc ii}}}

\newcommand{\oii}{\mbox{[O$\,${\sc ii}]}}
\newcommand{\oiii}{\mbox{[O$\,${\sc iii}]}}
\newcommand{\neiii}{\mbox{[Ne$\,${\sc iii}]}}
\newcommand{\oi}{\mbox{[O$\,${\sc i}]}}
\newcommand{\kms}{\mbox{${\;{\rm km\,s^{-1}}}$}}
\newcommand{\radm}{\mbox{$\;\mathrm{rad}\;\mathrm{m^{-2}}$}}
\newbox\grsign \setbox\grsign=\hbox{$>$}
\newdimen\grdimen \grdimen=\ht\grsign
\newbox\laxbox \newbox\gaxbox
\setbox\gaxbox=\hbox{\raise.5ex\hbox{$>$}\llap
        {\lower.5ex\hbox{$\sim$}}}\ht1=\grdimen\dp1=0pt
\setbox\laxbox=\hbox{\raise.5ex\hbox{$<$}\llap
        {\lower.5ex\hbox{$\sim$}}}\ht2=\grdimen\dp2=0pt
\def\clap#1{\hbox to 0pt{\hss #1\hss}}

\psfigurepath{/home-theory/helenj/SCR/Figures/B1221-423}

\title[The extraordinary radio galaxy MRC~B1221$-$423]{The
extraordinary radio galaxy MRC~B1221$-$423: probing deeper at radio
and optical wavelengths}
\author[H. M. Johnston et al.]{Helen M.
  Johnston,$^1$\thanks{E-mail: H.Johnston@physics.usyd.edu.au}
  Jess W. Broderick$^{1,2}$, Garret Cotter$^{3}$, Raffaella Morganti$^{4,5}$, \ 
  \newauthor
  and Richard W. Hunstead$^1$\\
  $^1$Sydney Institute for Astronomy, School of Physics, University of
  Sydney, NSW 2006, Australia \\
  $^2$School of Physics and Astronomy, University of Southampton,
  Southampton SO17 1BJ, UK\\
  $^3$Department of Astrophysics, University of Oxford, Keble Road,
  Oxford OX1 3RH, UK \\
  $^4$Netherlands Foundation for Research in Astronomy, Postbus 2,
  7990 AA Dwingeloo, the Netherlands \\
  $^5$Kapteyn Astronomical Institute, University of Groningen, P.O. Box 
  800, 9700 AV Groningen, the Netherlands} 
  \date{Received: }

\begin{document}

\maketitle

\begin{abstract}
     We present optical spectra and high-resolution multi-wavelength
     radio observations of the compact steep-spectrum radio source
     MRC~B1221$-$423 ($z=0.1706$).  MRC~B1221$-$423 is a very young
     ($\sim 10^5$~yr), powerful radio source which is undergoing a
     tidal interaction with a companion galaxy.  We find strong
     evidence of interaction between the AGN and its environment. The
     radio morphology is highly distorted, showing a dramatic
     interaction between the radio jet and the host galaxy, with the
     jet being turned almost back on itself.  \hi\ observations show
     strong absorption against the nucleus at an infall velocity of
     $\sim 250\,$\kms\ compared to the stellar velocity, as well as a
     second, broader component which may represent gas falling into
     the nucleus.  Optical spectra show that star formation is taking
     place across the whole system. Broad optical emission lines in
     the nucleus show evidence of outflow. Our observations confirm
     that MRC~B1221$-$423 is a young radio source in a gas-rich
     nuclear environment, and that there was a time delay of a few
     $\times 100$~Myr between the onset of star formation and the
     triggering of the AGN.

\end{abstract}


\section{Introduction}
\label{sec:intro}

There is growing evidence of strong links between the triggering of
powerful radio sources and gravitational interactions between
galaxies.  Many radio galaxies show signatures of tidal interactions,
such as tails, bridges, shells, and double nuclei
\citep[e.g.][]{hsb+86,bh92}.  These tidal encounters can lead to large
increases in the amount of material being fed to the central black
hole, thereby triggering the strong radio emission.  The same tidal
event that dumps gas in the vicinity of the nuclear black hole can
produce a burst of star formation, and hence a population of young
stars. 

Further, the mechanical energy associated with relativistic radio jets
can have a profound effect on the formation and evolution of galaxies,
by heating the interstellar gas to the point where stars can no longer
form. This \emph{radio-mode feedback} has been suggested as the
mechanism which suppresses star formation in massive galaxies; without
such a feedback mechanism, galaxy evolution models do not match the
observed properties of galaxies \citep{csw+06}.

So both starbursts and the growth of the nuclear black hole are
intimately associated with the merger history of the galaxy.  However,
key questions remain unanswered. If mergers are responsible for the
formation of both AGN and starbursts, why do we sometimes see one
without the other? When we do see both, we usually see a young radio
source together with an older starburst, suggesting that there is a
delay between the starburst and the onset of AGN activity.  What is
the reason for this delay?

The compact steep-spectrum (CSS) sources are ideal laboratories for
testing these connections.  CSS sources are powerful but compact radio
sources, with sizes 1--20~kpc, so that the radio source lies well
within its host galaxy. Their radio spectra are steep, $\alpha <
-0.5$\ ($S_\nu \propto \nu^\alpha$), peaking below 500~MHz.  They have
the same distribution in redshift as the rest of the powerful radio
source population \citep[see][for a review]{ode98}.  The weight of
evidence now is that CSS sources are small because they are young; if
fuelling is sustained they are expected to evolve into classical large
double-lobed radio sources \citep{ode98}. All CSS sources accessible
to detailed optical investigations show evidence for recent
interactions and/or mergers, and evidence for young stellar
populations and galaxy-wide starbursts have been found in many of
these objects \citep[see][for a review]{hol09}.

\begin{figure}
     \centerline{\psfig{figure=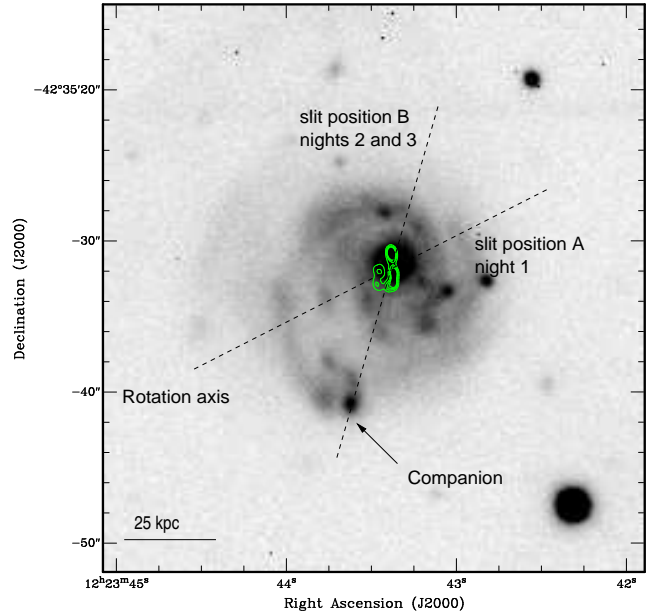,width=\columnwidth,clip=t}}
     \caption{$V$-band NTT image of the host galaxy
     \citep[from][]{jhcs05}, showing knotty
       features and tidal tails, strongly suggesting an ongoing
       interaction with the companion galaxy (arrowed). The 12~mm
       radio contours (Fig.~\protect{\ref{fig:contour_plot}}) are
       overlaid to show the scale of the radio source in relation to
       the galaxy. The two slit positions
       (Table~\protect{\ref{tab:obs}}) are shown; spectrum A was taken
       along the rotation axis of the galaxy (see
       \S~\ref{sec:Observ-data-reduct}), while spectrum B was taken
       along the line joining the galaxy to its companion, which is
       close to the axis of the radio source.}
     \label{fig:slitpos}
\end{figure}

MRC~B1221$-$423 is a member of a sample of southern radio sources
similar in power to 3C sources \citep{bh06}, with a radio power of
$P_{\mathrm{1.4 GHz}} = 1.8 \times 10^{26}\;{\rm W\,Hz}^{-1}$. The
optical emission of the host galaxy is distorted, showing clear
signatures of interaction, including tidal tails and shells
\citep{shp03}.  Multicolour images show knotty blue star-forming
regions, while the likely merging galaxy to the south is also
significantly bluer than the host galaxy (Johnston et al. 2005; see
also Figure~\ref{fig:slitpos})\nocite{jhcs05}. With a redshift of
$z=0.1706$\ \citep{scrw93}, MRC~B1221$-$423 is one of the nearest CSS
sources; in the 3C sample, we need to go to $z=0.3$\ to find the first
compact source as powerful as MRC~B1221$-$423.  The radio source has
double lobes which lie well within the envelope of the host galaxy
\citep{shp03}.  The age of the radio source, estimated from the power
of the radio jets and total stored energy in the synchrotron plasma of
the lobes \citep{rs91}, is only $\sim 10^5$~yr. As a young, nearby,
bright CSS source, MRC~B1221$-$423 is therefore an ideal candidate for
testing the relationship between the AGN and its environment.

The goal of this study is to understand the relationship between the
central AGN, which is the observational signpost of a growing black
hole, and the environment of the galaxy. MRC~B1221$-$423 provides a
unique opportunity to measure both the age of the radio source and the
star formation history of the galaxy, both of which are connected to
the merger history of the galaxy.

In a previous paper (\citealt{jhcs05}, henceforth Paper I) we
presented an analysis of multi-colour \emph{BVRIK} images of
MRC~B1221$-$423, which we used to analyse the change in stellar
populations across the galaxy and its companion. We found evidence for
three distinct stellar populations, with ages of about 15~Gyr, 300~Myr
and less than 10~Myr. We concluded that the interaction with the
companion galaxy triggered star formation in both galaxies, and after
a substantial time delay, the infalling gas triggered activity in the
central supermassive black hole.

In order to better understand the connection between the radio source
and its environment, we have undertaken further radio and optical
observations of MRC~B1221$-$423, which we present in this paper. In
sections~\ref{sec:Radio-imaging} and~\ref{sec:Radio-spectroscopy} we
describe the radio images and \hi\ absorption spectra of
MRC~B1221$-$423, and in section~\ref{sec:Optical-observations} we
describe the optical spectra. We have assumed a flat $\Lambda$-CDM
cosmology with $H_0 = 71\;{\rm km\,s^{-1}\,Mpc^{-1}}$,
$\Omega_\mathrm{M}=0.27$ and $\Omega_\mathrm{\Lambda}=0.73$\
throughout. At the distance of the galaxy, 1~arcsec corresponds to a
distance of 2.876~kpc.

\section{Radio imaging}
\label{sec:Radio-imaging}


We observed MRC~B1221$-$423 at 1344, 2282 and 18\,880.5 $+$
19\,648.5~MHz (20~cm, 13~cm and 12~mm wavebands) with the Australia
Telescope Compact Array (ATCA), and also extracted unpublished,
archival data at 1384 and 2496 MHz from the ATCA online
archive\footnote{http://atoa.atnf.csiro.au}.
Table~\ref{tab:radio_obslog} shows the details of these
observations. 

The data were reduced using standard procedures in {\sc miriad}
\citep{stw95}. The effective bandwidth is either 104 MHz (13 channels,
each of width 8 MHz) or 56 MHz (14 channels, each of width 4
MHz). However, at both 1384 and 2282 MHz, not all of the band is
usable because the ATCA suffers from self-generated interference at
integer multiples of 128 MHz. Hence, there is a slight shift of the
effective frequency in both cases (see Table~\ref{tab:radio_obslog}).

PKS B1934$-$638 was the primary calibrator at 20 and 13~cm, except for
the observations from 2003 November where it was not observed. In this
case, PKS B0823$-$500 as well as the 2280 MHz data from 2003 August
were used to determine the flux density scale; we estimate that the
uncertainty in this procedure is a few per cent at most. At 18.9 and
19.6 GHz, flux calibration was carried out using observations of
Jupiter; we used the shortest baseline only, where Jupiter is not
significantly resolved out.

The deconvolution and imaging procedure at each frequency consisted of
a number of iterations of {\sc clean} plus self-calibration. Only
after several rounds of phase-only self-calibration was a full
amplitude and phase solution attempted. To ensure that the source
model derived from the {\sc clean} components was as simple as
possible (especially in the 12-mm waveband), natural weighting was
used while self-calibrating. When it was clear that there was no
further improvement in the image quality, both naturally- and
uniformly-weighted images were created from the final self-calibrated
visibilities. Table~\ref{tab:radio_obslog} contains a summary of the
angular resolution and RMS noise level in the vicinity of
MRC~B1221$-$423 in each total intensity image. The 18.9 and 19.6 GHz
images were restored using the same beam size (that of the 18.9 GHz
data) to allow these images to be averaged together for further
analysis (average frequency 19\,264.5 MHz). We estimate that the
astrometry in the millimetre-wavelength images is accurate to
$\sim$0.1 arcsec.

\begin{table*}
\caption{Log of ATCA radio observations.}
\label{tab:radio_obslog}
\begin{center}
\begin{tabular}{@{}lcccccccccc}
\hline
     &               & Observing & Effective & Integration &
\multicolumn{2}{c}{Angular resolution$^{a}$} & \multicolumn{2}{c}{Beam} & \multicolumn{2}{c}{RMS noise$^{c}$} 
 \\
Date &         & frequency & bandwidth & time &
\multicolumn{2}{c}{(arcsec $\times$ arcsec)}  & \multicolumn{2}{c}{PA$^b$} & \multicolumn{2}{c}{(mJy beam$^{-1}$)}
  \\
(UT) & Array & (MHz)     & (MHz)     & (h) &
N & U & N & U & N & U \\
\hline
2004 Jan 9  & 6A & 1344 & 104 & 8.44 &
$14.05 \times 9.58$ & $8.12 \times 5.71$ & $-1\fdg1$ & $-0\fdg 3$ & 0.45 & 0.55 \\
1997 Jul 7 & 6A & 1384 & 104\clap{$^{\:\,d}$}& 0.06 & 
$13.48 \times 10.02$ & $10.12 \times 6.60$  & $-24\fdg 3$ & $-24\fdg 8$ & 1.2 & 1.4 \\
2003 Aug 23  & EW367 & 2282 & 56\clap{$^{\:\,e}$} & 3.37 & 
\multirow{2}{*}{$14.19 \times 13.64$} & \multirow{2}{*}{$4.20 \times 3.31$} & 
\multirow{2}{*}{$16\fdg 0$} & \multirow{2}{*}{$16\fdg 4$} & \multirow{2}{*}{0.42} & \multirow{2}{*}{0.55} \\ 
2003 Nov 22  & 1.5D & 2282 & 56\clap{$^{\:\,e}$} & 6.18 & \\
1997 Jul 7 & 6A & 2496 & 104 & 0.06 & 
$7.45 \times 5.60$ & $5.33 \times 3.17$ & $-24\fdg 4$ & $-23\fdg 9$ & 1.3 & 1.8 \\
\multirow{2}{*}{2003 Jul 27} & \multirow{2}{*}{6D} & 18\,880.5 &
104 & 7.59 & 
$1.09 \times 0.67$ & $0.57 \times 0.34$ & $9\fdg 6$ & $8\fdg 0$ & 0.11 & 0.18 \\
 & & 19\,648.5 & 104 & 7.59 & 
$1.09 \times 0.67$ & $0.57 \times 0.34$ & $9\fdg 6$ & $8\fdg 0$ & 0.11 & 0.18 \\
\hline
\end{tabular}

\medskip
\begin{minipage}{\textwidth}
$^a$N -- natural weighting; U -- uniform weighting.

$^b$Measured north through east

$^c$RMS noise level in the vicinity of MRC B1221$-$423 in the total
intensity image

$^d$One of the 8 MHz channels is flagged due to self-interference at
1408 MHz. The effective frequency is shifted slightly to 1382 MHz.

$^e$One of the 4 MHz channels is flagged due to self-interference at
2304 MHz. The effective frequency is shifted slightly to 2280 MHz.
\end{minipage}
\end{center}
\end{table*}

\subsection{Morphology}\label{section morphology}

\begin{figure}
  \centerline{
    \psfig{file=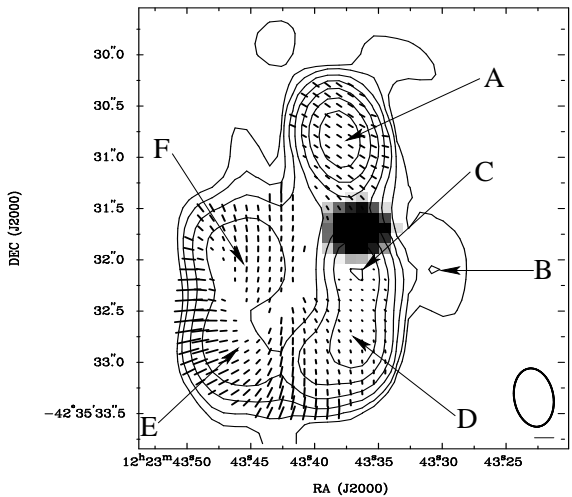,width=7.5cm} 
  }

 \caption{Total-intensity, uniformly-weighted contour map of
   MRC~B1221$-$423 at 19.3 GHz, with the position of the optical
   nucleus shown in greyscale. The lowest contour is 0.75 mJy
   beam$^{-1}$, with each successive contour corresponding to a
   doubling of the surface brightness. Vectors show the projected
   electric field at each pixel location (not corrected for Faraday
   rotation; pixel size $0.1~\mathrm{arcsec} \times
   0.1~\mathrm{arcsec}$); the scale bar in the bottom right-hand
   corner corresponds to 50 per cent fractional linear polarisation. A
   S/N cutoff of 5 in polarised intensity was used when plotting the
   polarisation vectors. The various components have been
   labelled. The synthesised beam is shown in the bottom right-hand
   corner.}
  \label{fig:contour_plot}
\end{figure}

At 8640 MHz, MRC~B1221$-$423 is resolved into a north-south double
structure with a projected separation of 4.3 kpc (1.5 arcsec), as well
as a third component to the east \citep{shp03}. However, the finer
angular resolution at 19.3~GHz reveals that MRC~B1221$-$423 has a
particularly striking radio morphology akin to a string of pearls
(Figure~\ref{fig:contour_plot}). The southern jet has undergone a
dramatic interaction with the host galaxy, being bent through a full
180\degr. The projected length of the jet from the position of the
optical nucleus to the peak of the furthest knot (knot F) is
8.6 kpc. On the other hand, the northern jet is much shorter: the
projected separation is only 2.5 kpc.

\subsection{Spectral energy distribution}\label{sec:sed}

\begin{figure}
\centerline{\psfig{file=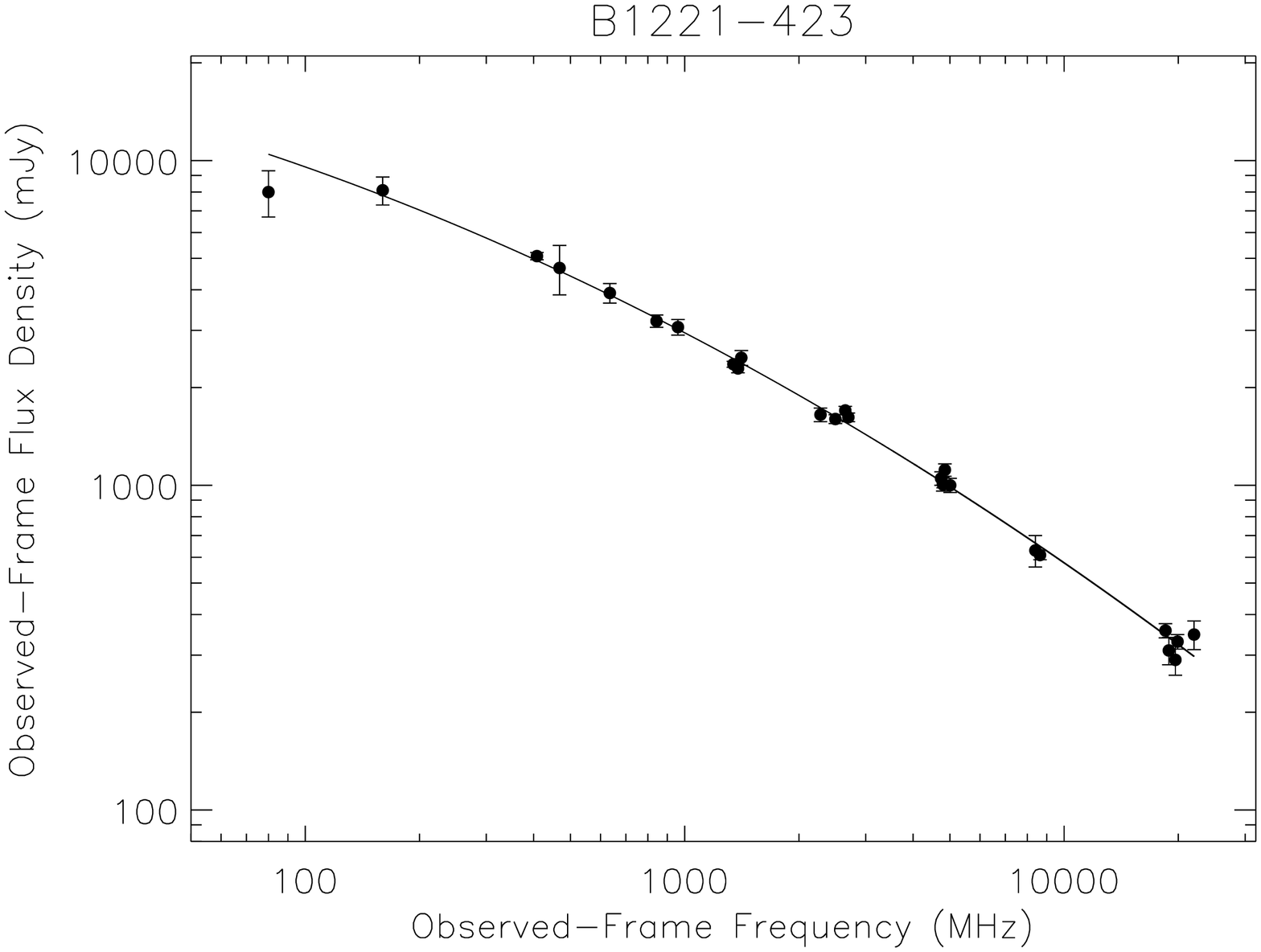,width=\columnwidth,angle=90,clip=y}}
 \caption{The observed-frame radio SED for MRC~B1221$-$423, using all of
   the flux densities from Table~\ref{table:radio_fluxes}. The 843,
   4800 and 4850 MHz data points are the non-weighted averages of the
   two individual measurements at each of these frequencies. A
   second-order polynomial has been fitted to the SED: $\ln(S_{\nu}) =
   10.15 - 0.01619 \ln (\nu) - 0.04295 [\ln (\nu)]^2$ with $\nu$ in MHz
   and $S_{\nu}$ in mJy. Though nearly all of the Parkes flux
   densities are affected by blending, the fit to the SED is not
   significantly altered if these data points are excluded from the
   fit (the two fits agree typically to within a few per cent).}
\label{fig:radio_SED}
\end{figure}

Figure~\ref{fig:radio_SED} shows the spectral energy distribution of
MRC~B1221$-$423; the fluxes are tabulated in
Table~\ref{table:radio_fluxes}. Because the source is only slightly
extended at 20 and 13 cm, the 1344, 1382, 2280 and 2496 MHz flux
densities were each determined by fitting an elliptical Gaussian with
the {\sc miriad} task {\sc imfit}.  However, it was found that the
most accurate way to measure the flux density at millimetre
wavelengths was to sum the flux in a polygon that enclosed the
source. The uncertainty for each of the new ATCA measurements is
dominated by the internal calibration uncertainty.  While the 20 and
13 cm flux densities are not affected by the weighting scheme used, we
have quoted the 18.9 and 19.6~GHz flux densities from the
naturally-weighted images, which are approximately three per cent
higher than the values from the uniformly-weighted images. This will
be due to some of the extended emission being resolved in the
uniformly-weighted images because of the finer resolution. However,
the naturally-weighted images themselves may be missing some extended
flux, as, for example, can be seen by comparing our results with the
$\sim$20 GHz flux densities reported by \citet{rpg+06} and
\citet{mse+10}, which were derived using lower-resolution ATCA data
than ours.

The observed-frame radio spectral energy distribution (SED) between 80
MHz and 22 GHz steepens with increasing frequency
(Figure~\ref{fig:radio_SED}). The spectral index $\alpha$\ varies from
$-0.53$ at 408 MHz to $-0.88$ at 22 GHz. Moreover, the 1.4 GHz
luminosity is $1.8 \times 10^{26}$ W Hz$^{-1}$ for our assumed
cosmological model. Further low-frequency measurements are needed to
determine accurately the turnover frequency.



\subsection{Linear polarisation}\label{section frac pol}

\citet{shp03} investigated the polarimetric properties of MRC~B1221$-$423
at 4800 and 8640 MHz, finding that that the fractional linear
polarisation $m$ is as high as 35 per cent at 8640 MHz (resolution 1.4
arcsec $\times$ 0.63 arcsec). We now extend this analysis with our new
ATCA high-frequency data. In Figure~\ref{fig:contour_plot}, we have
plotted polarisation vectors at 19.3 GHz (resolution 0.57 arcsec
$\times$ 0.34 arcsec). The vectors themselves were computed by first
forming Stokes $Q$ and $U$ images at both 18.9 and 19.6 GHz from the
self-calibrated visibilities; these images were restored with the same
synthesised beam size. Averaged images were then produced for Stokes
$Q$ and $U$ at 19.3 GHz, from which a total polarised intensity ($P =
\sqrt{Q^2+U^2}$) map, corrected for Ricean bias, was also formed.  The
average RMS noise level in the 19.3 GHz Stokes $Q$ and $U$ images
($\sigma_{QU}$) in the vicinity of MRC~B1221$-$423 is about 0.1 mJy
beam$^{-1}$. Note that bandwidth depolarisation is negligible for
MRC~B1221$-$423 in the 18.9 and 19.6 GHz bands.

High fractional linear polarisation is also observed at millimetre
wavelengths, particularly at the eastern edge of the source, where the
mean fractional polarisation is $\sim$50 per cent, with some pixels
having $m \ga 60$ per cent. The fractional polarisation is also
significant at the southern edge of the source (30--40 per cent), as
well as between knots D and E ($\sim$25 per cent on average). Over the
entire source, the median fractional polarisation of the pixels is about
14 per cent. Further polarisation statistics are shown in
Table~\ref{table: radio pol}.

\begin{table}
  \caption{19.3 GHz flux density, fractional polarisation and
      polarisation position angle measurements in the
      uniformly-weighted image
      (Fig.~\protect{\ref{fig:contour_plot}}). The measurements were
      made at the position of the peak pixel in Stokes $I$ for each of
      the source components.}
  \label{table: radio pol}
\begin{tabular}{cccc}
\hline
\multicolumn{1}{c}{Component} & $S_I$ & \multicolumn{1}{c}{$m$} & \multicolumn{1}{c}{$\chi$}   \\
& \multicolumn{1}{c}{(mJy beam$^{-1}$)} & \multicolumn{1}{c}{(\%)} & \multicolumn{1}{c}{(\degr)}  \\
\hline
A & 89.9 & $13.0 \pm 0.2$ & $52.1 \pm 0.3  $   \\
B & 1.5 & $< 20$\rlap{$^a$} & $\cdots$ \\
C & 48.4 & $1.9 \pm 0.2$ & $59 \pm 3$ \\
D & 42.0 & $5.7 \pm 0.3$ & $14.3 \pm 1.2$ \\
E & 11.8 & $14.7 \pm 0.9$ & $-67.0 \pm 1.6$  \\
F & 11.9 & $14.7 \pm 0.9$ & $-1.3 \pm 1.6$ \\
\hline
\end{tabular}

$^a$3$\sigma$ upper limit. \\
\end{table}

The orientation of the projected magnetic field with respect to the
jet path varies significantly across the source. Knots A and C have
similar polarisation position angles even though there is a
significant difference in the degree of polarisation; in these regions
the magnetic field is neither parallel nor perpendicular to the jet
path. While the magnetic field becomes increasingly perpendicular to
the jet path at knot D, it then becomes strikingly parallel to the jet
between knots D and F where the extreme bending occurs. The magnetic
field is also oriented circumferentially along the outer contours in
this latter region, which was also observed in the lower-resolution
8640 MHz data \citep[][]{shp03}. In complete contrast, knot F exhibits
magnetic fields that are closely perpendicular to the jet path.

Note that after correcting for Faraday rotation (see below), the
intrinsic and observed polarisation position angles are estimated to
agree to within $\la 5^{\circ}$\ except for knot A, where the
vectors need to be rotated anti-clockwise by $\sim$10--15$^{\circ}$.

\subsection{Rotation measure synthesis}\label{sect:RM}

Faraday rotation measures (RMs) provide an important insight into the
properties of the magnetoionic environment along the line of sight
towards a radio source. Traditionally, the RM is determined from a
linear fit between the polarisation position angle $\chi$ and
$\lambda^2$. However, this method suffers from several problems, such
as position angle ambiguities. A superior Fourier-based technique,
known as Faraday rotation measure synthesis, overcomes the problems
associated with $\chi$--$\lambda^2$ fitting. Detailed descriptions of
the method can be found in \citet{bb05} and \citet{hbe09}. In brief,
the complex linear polarisation $P(\lambda^2) = Q(\lambda^2) +
iU(\lambda^2)$ and complex Faraday dispersion function $F(\phi)$ are
related via a relationship that is very similar to a Fourier
transform. The Faraday depth $\phi$, which is a more general quantity
than the RM, is defined as 
\begin{equation}\label{eq far depth} \phi = 0.812 \int^{L}_{0} n_{\rm e} \bmath{B} \cdot d\bmath{l} 
\end{equation}
where $L$ is the path length through the magnetised plasma in pc,
$n_{\rm e}$ is the thermal electron density in cm$^{-3}$, and
$\bmath{B}$ is the magnetic field in $\mu$G. Measurements of $P$ in a
number of channels across one or more bandpasses can thus be used to
form an `RM data cube'; a slice through this cube at a particular
spatial position gives a Faraday depth spectrum ($|F(\phi)|$ versus
$\phi$).

\subsubsection{Low-frequency RM synthesis}

First, we performed RM synthesis at 20 and 13 cm using three different
approaches: 1344 MHz alone ($13 \times 8$~MHz channels), 2280 MHz
alone ($13 \times 4$~MHz channels) and by combining the 1344 and 2280
MHz bands. There is no significant advantage in also including the
1382 and 2496 MHz observations because $\sigma_{QU}$ is much higher
than at 1344 and 2280 MHz ($\sim$0.12 and $\sim$0.20 mJy beam$^{-1}$,
respectively, in the two latter cases).  Each Stokes $Q$ and $U$
individual channel image was restored at the resolution of the 1344
MHz data ($8.12$ arcsec $\times$ $5.71$ arcsec; PA $= -0 \fdg
3$). Because MRC~B1221$-$423 is only slightly extended at these
frequencies, we formed Faraday depth spectra at the position of peak
intensity in Stokes $I$. Though we cannot spatially resolve the
individual components of the source, RM synthesis is still useful: the
$\lambda^2$ coverage is greater than at higher frequencies, which
reduces the uncertainty in the Faraday depth because the resolution in
$\phi$-space is much higher. Therefore multiple Faraday depth
components can potentially be resolved in $\phi$-space even though we
cannot spatially distinguish the different Faraday screens.

Inverse-variance weighting was applied to each channel when forming
the Faraday depth spectra. A spectral index correction was also
implemented \citep[see][]{bb05}.  The spectra were {\sc clean}ed to
remove the sidelobes that are present due to incomplete $\lambda^2$
coverage; the response function is referred to as the rotation measure
spread function (RMSF). A Gaussian `restoring beam' was used when
constructing each {\sc clean}ed spectrum \citep[see][for further
details]{hbe09}; the resolutions in $\phi$-space are 591, 5123 and 74
rad m$^{-2}$ FWHM at 1344 MHz alone, 2280 MHz alone, and 1344 $+$ 2280
MHz combined, respectively.

Both the 1344 and 2280 MHz spectra are found to consist only of a
single point-like Faraday depth component. However, the component
shifts position between the two frequencies: the position of the peak
is $-35 \pm 4 \mathrm{rad}\;\mathrm{m}^{-2}$ at 1344 MHz, but $-220
\pm 30$ rad m$^{-2}$ at 2280 MHz. The uncertainties take into account
both the noise statistics and the estimated residual instrumental
polarisation. The components are bright enough such that the Ricean
bias is not significant.

\begin{figure}
  \centerline{\psfig{file=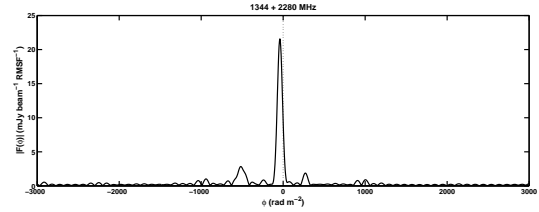,width=\columnwidth}}
  \caption{{\sc clean}ed 1344 $+$ 2280 MHz Faraday depth spectrum for
  MRC~B1221$-$423 at the position of peak intensity in Stokes $I$. The
  dotted line indicates a Faraday depth of zero. The Ricean bias has
  not been removed. The most prominent peak can be associated with the
  foreground Galactic RM.}
\label{fig:RM_synthesis1}
\end{figure}

The shift in the position of the peak suggests that there is an
unresolved distribution of components whose relative strengths vary
with frequency. Indeed, multiple components are found when the 1344
and 2280 MHz data are combined to form a higher-resolution spectrum
(Figure~\ref{fig:RM_synthesis1}). Apart from the bright peak at $-40
\pm 2 \mathrm{rad}\;\mathrm{m}^{-2}$, there are weaker components
above 10$\sigma$ (about 1.3 mJy beam$^{-1}$ RMSF$^{-1}$) at $+270$\
and $-510$ rad m$^{-2}$, the latter being extended over 1.6
beamdwidths. However, these results must be interpreted with caution:
the gap between the 20 and 13 cm bands introduces large sidelobes into
the RMSF, which complicates the deconvolution process (though see the
results at higher frequency below).

The Faraday depth of the brightest component in
Figure~\ref{fig:RM_synthesis1} is very similar to the mean RM of $-35$
rad m$^{-2}$ calculated by \citet{shp03} using the traditional linear
fitting approach at 4800 $+$ 8640 MHz. As discussed by \citet{shp03},
this result is consistent with the Galactic foreground RM in this
region of the sky. We were also able to construct Faraday depth
spectra at 1344 MHz for two nearby sources in the field: SUMSS
J122328$-$423436 and SUMSS J122316$-$423922. Again, the derived
Faraday depths are consistent with a Galactic magnetoionic screen
being a significant, if not the main source of Faraday rotation.

\subsubsection{High-frequency RM synthesis}

We also attempted RM synthesis at higher frequencies by combining
re-analysed 8640 MHz data (uniform weighting; resolution 1.14 arcsec
$\times$ 0.72 arcsec; PA $= -9.4^{\circ}$) with our
millimetre-wavelength observations (naturally weighted but restored at
the resolution of the 8640 MHz data); there are a total of 39
channels, each of width 8 MHz. Though the angular resolution is
superior at these frequencies, the $\lambda^2$ coverage is poorer:
the FWHM in $\phi$-space is about 2600 rad m$^{-2}$ for
inverse-variance weighting, and, like at 1344 $+$ 2280 MHz, the RMSF
contains large sidelobes because of the gaps between the
bands. Because it is very easy for the peak of $|F(\phi)|$ to be
shifted to the position of a nearby sidelobe for the lower S/N pixels,
we only measured Faraday depths for the pixels with peak polarised
flux densities greater than 1 mJy beam$^{-1}$ RMSF$^{-1}$, after
applying a spectral index correction as above.

\begin{figure}
  \centerline{\psfig{file=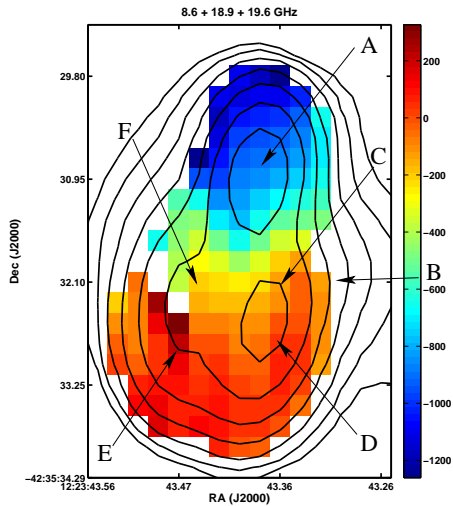,width=6.5cm}}
  \caption{8.6 $+$ 18.9 $+$ 19.6 GHz observed-frame Faraday depth map
    for MRC~B1221$-$423 (angular resolution resolution 1.14 arcsec $\times$
    0.72 arcsec; beam PA $-9.4^{\circ}$; pixel size 0.23 arcsec $\times$
    0.23 arcsec). The total-intensity radio contours are at 8640~MHz;
    the lowest contour is 2 mJy beam$^{-1}$ and each successive
    contour corresponds to a doubling of the surface brightness. The
    colour bar has units of $\radm$. The typical uncertainty in the
    Faraday depth for a given pixel is about $30\radm$. The letters
    label the positions of the components seen in the 19.3~GHz map
    (Fig.~\protect{\ref{fig:contour_plot}). }
  }
\label{fig:RM_synthesis2}
\end{figure}

Figure~\ref{fig:RM_synthesis2} shows a $3\,\mathrm{cm} +
12\,\mathrm{mm}$\ Faraday depth map for MRC~B1221$-$423. Each pixel
has only one $\phi$ component (though the sidelobes significantly
complicate the search for secondary components), and none of these
components are extended. The Faraday depth varies from $-1260$ to
$+330\; \mathrm{rad}\;\mathrm{m}^{-2}$, with a steep gradient from
north to south: the mean Faraday depth for the top half of the source
is $\sim -800$ rad m$^{-2}$, while the mean for the bottom half is
$\sim -30$ rad m$^{-2}$. The latter value is close to the Galactic
foreground RM (see above). The steep gradient in $\phi$ strongly
suggests that the very large Faraday depths associated with the
northern lobe are a consequence of a magnetoionic screen that is local
to the source. The rest-frame Faraday depth will be a factor of
$(1+z)^2 = 1.37$\ times higher than the observed-frame value after the
Galactic contribution is removed from the latter. Assuming that the
foreground Faraday depth is $\sim -40
\;\mathrm{rad}\;\mathrm{m}^{-2}$, then the rest-frame Faraday depth at
the position of peak intensity in Stokes $I$ for the northern lobe is
about $-1100 \;\mathrm{rad}\;\mathrm{m}^{-2}$. From Equation~\ref{eq
far depth}, large values of phi are a consequence of one or more of
the following along the line of sight: (1) a dense medium, (2) a
highly magnetised medium, and (3) a long path length. The radio
observations alone are not sufficient to disentangle the relative
contributions.

\begin{figure}
\centerline{\psfig{file=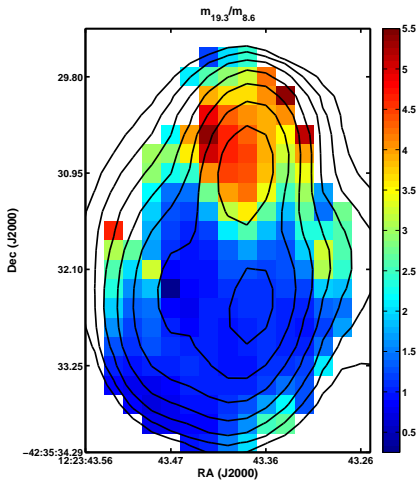,width=6.5cm}}
\caption{Depolarisation ratio map between 8.6 and 19.3 GHz. The radio
   contour scheme is the same as in Figure~\ref{fig:RM_synthesis2}. A
   S/N cutoff of 5 in polarised intensity was used at both 8.6 and
   19.3 GHz when constructing the map.}
\label{fig:depol_map}
\end{figure}

To investigate further the magnetoionic environment of
MRC~B1221$-$423, we have also constructed a map of the depolarisation
ratio $m_{19264.5}/m_{8640}$ (Figure~\ref{fig:depol_map}). The
northern lobe is significantly depolarised with decreasing frequency,
implying small-scale inhomogeneities in the Faraday screen. This is
likely to explain why there is not a prominent $\phi$ component at
$\sim -1000 \;\mathrm{rad}\;\mathrm{m}^{-2}$ in
Figure~\ref{fig:RM_synthesis1}. On the other hand, the southern jet is
not significantly depolarised between 19264.5 and 8640 MHz; the
fractional polarisation in fact increases with decreasing frequency in
some areas.

It is possible that we are seeing an orientation effect, with the
radio source fore-shortened and the northern lobe seen through the
galaxy. We discuss this possibility in Section~\ref{sec:discussion}.

\section{Radio spectroscopy}
\label{sec:Radio-spectroscopy}

The \hi\ observations of MRC~B1221$-$423 were done with the ATCA for
$2 \times 12$~hrs on 2004 January 9 and 2004 July 17 using the 6-km
array-configuration.  All observations were done with 16 MHz bandwidth
and a central frequency (for the first IF) set to 1214 MHz. The first
observation used 256 channels and two IFs while the second observation
used 512 channels and only one IF.

We used PKS~B1934$-$638 as bandpass and flux calibrator. To monitor
the system gain and phase changes, PKS~B1215$-$457 was observed as a
secondary calibrator for ten minutes every hour.

For the data reduction and visualisation we used the {\sc miriad} and
{\sc karma} \citep{goo96} data reduction packages. For the line
observations, after flagging and calibration, we separated the
continuum from the line data in each individual data set by fitting
either a first or a second order polynomial to the line-free channels.

The continuum was subtracted using {\sc uvlin} by making a
second-order fit to the line-free channels of each visibility record
and subtracting this fit from the spectrum. After cleaning, Hanning
smoothing was applied to each line cube.  The final cube was derived
by combining the datasets for the two observations. The beam size in
the resulting uniformly weighted cube is $9\farcs6 \times 6\farcs4$,
$\mathrm{PA} =-1\degr$. The rms noise in the Hanning cube is
$0.77\;\mathrm{mJy}\;\mathrm{beam}^{-1}$ per channel. Velocity
resolution is 26 \kms\ after Hanning.

\begin{figure}
     \centerline{\psfig{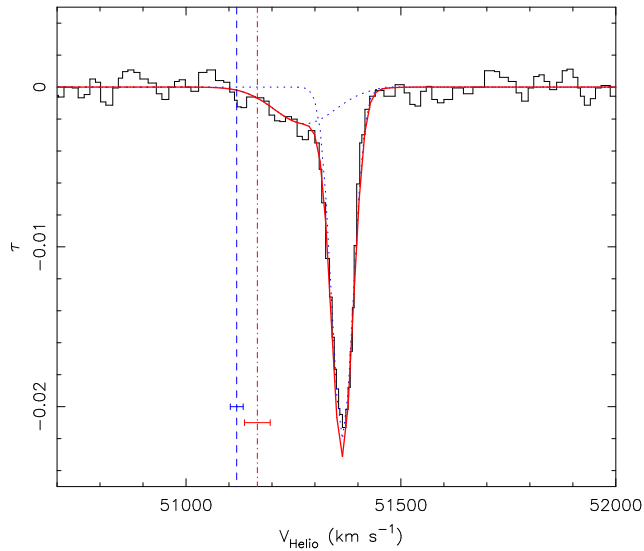}}
     \caption{\hi\ absorption against the nucleus of MRC~B1221$-$423,
       showing large amounts of redshifted (infalling) material
       compared to the systemic velocity (dashed line). There is also
       a broad wing of absorption extending blueward of the narrow
       absorption component. The thick red line shows the sum of two
       gaussians fitted to the profile; the individual components are
       shown as the dotted blue lines. The vertical dashed line with
       error bar shows the stellar velocity ($cz = 51118 \pm 15$\kms,
       see \S~\ref{sec:Velocities}), while the vertical dash-dotted
       line shows the mean emission-line velocity ($cz= 51166 \pm
       30$\kms, see \S~\ref{sec:Emission-line-gas}).  }
   \label{fig:HIprofile}
\end{figure}

\hi\ in absorption was clearly detected. The absorption, shown in
Figure~\ref{fig:HIprofile}, consists of a deep and relatively narrow
component with a broad wing.  The narrow component (FWHM $\sim 90$
\kms) has a peak of $-40.5$~mJy centred on $\sim 51370$ \kms.  The
optical depth of this component is $\tau \sim 0.02$.  Compared to the
systemic velocity derived from the stars ($51118 \pm 15\kms$; see
\S~\ref{sec:Stellar-continuum}), this absorption component is
redshifted by $\sim 250$ \kms.  The broad wing (estimated full-width
at zero intensity (FWZI) $\sim 400$\kms) is blueshifted with respect
to the narrow component by $\sim 150$\kms. It covers a velocity range
extending to $\sim 51100$\kms, therefore reaching velocities slightly
blueshifted compared to the systemic velocity of the galaxy. The peak
absorption of this wing has an optical depth of $\tau \sim 0.003$ (see
Figure~\ref{fig:HIprofile}).

No blueshifted component (which could be connected with outflowing
gas) was detected, to an optical depth limit of $\tau \sim 0.001$
($3\sigma$). This is comparable with the optical depth detected in
broad absorption components associated with outflows in other radio
galaxies \citep{mto05}, so we would have been able to detect such
outflows if they were present.

\section{Optical observations}
\label{sec:Optical-observations}

\subsection{Observations and data reduction}
\label{sec:Observ-data-reduct}

Optical spectra were obtained on 2004 May 14--16 using the Double Beam
Spectrograph on the ANU 2.3-m telescope at Siding Spring Observatory
(Table~\ref{tab:obs}).  The detectors were two SITe $1752 \times 532$\
CCDs with 15-$\mu$m pixels. For the first night, a dichroic filter
with a cross-over wavelength of 6300$\,$\AA\ was used to split the
light into the two arms of the spectrograph. Two gratings with
$600\;\mathrm{grooves}\,\mathrm{mm}^{-1}$\ were used, giving a
wavelength resolution of $2.2\,\mathrm{\AA}$\ over the wavelength
range 4200--6120$\,$\AA\ in the blue and 6270--7130$\,$\AA\ in the
red. 

\citet{shp03} used spectra taken at three different position
angles to determine the angle of maximum gas rotation velocity; the
(projected) axis of rotation was assumed to be perpendicular to this.
The slit was oriented along this rotation axis, at a position angle of
118\degr (slit position A; Figure~\ref{fig:slitpos}). This set-up gave
us high-resolution spectra with minimum rotational broadening of the
line profiles.

For the remaining two nights, a plane mirror was used in place of the
dichroic, so all light was sent to the blue arm of the spectrograph. A
$300\;\mathrm{grooves}\,\mathrm{mm}^{-1}$\ grating gave us a
wavelength resolution of $4.3\,\mathrm{\AA}$\ over the range
3700--7500$\,$\AA.  The slit was oriented at a position angle of
$-18\degr$\ so as to place both the galaxy and its interacting
companion on the slit (slit position B; Fig.~\ref{fig:slitpos}). This
set-up gave us slightly lower wavelength resolution but continuous
wavelength coverage of the blue region (rest-frame 3160--6400$\,$\AA),
which is useful for age-dating stellar populations.

The spatial scale on the detector was 0\farcs91/pixel, corresponding
to a linear scale of 2.6$\,$kpc per pixel.  On all three nights, a
slit-width of 1\farcs5 was used, and pairs of $1800\,\mathrm{s}$\ 
exposures were bracketed with NeAr arc-lamp exposures.

\begin{table}
\caption{Journal of optical spectroscopic observations of
  MRC~B1221$-$423. Two different set-ups were used, on the first night
  and remaining nights respectively.  The columns show the date of the
  observation, the wavelength range and the resolution, the exposure
  time, the position angle of the slit on the sky, and the median
  seeing of the observations. The final column shows the slit position
  (Fig.~\protect{~\ref{fig:slitpos}}). }
\label{tab:obs}\addtolength{\tabcolsep}{-2pt}
\begin{tabular}{l r@{--}l crrcc }
\hline
UT date & \multicolumn{2}{c}{$\lambda$\ range} &
  FWHM   & \multicolumn{1}{c}{$t_\mathrm{exp}$} &
  \multicolumn{1}{c}{PA} & Seeing \\
        & \multicolumn{2}{c}{(\AA)} &
   (\AA) & \multicolumn{1}{c}{(ks)} & 
  \multicolumn{1}{c}{(\degr)} & (arcsec) \\
\hline
2004 May 14 & 4200 & 6120 & 2.2 & 14.4 &   118 & 1.4 & \\
            & 6450 & 8000 & 2.2 & 12.6 &   118 & 1.4 &
            \raisebox{1.5ex}[0pt]{A} \\
2004 May 15 & 3700 & 7500 & 4.3 & 23.4 & $-18$ & 1.6 & \\
2004 May 16 & 3700 & 7500 & 4.3 & 23.4 & $-18$ & 1.5 &
            \raisebox{1.5ex}[0pt]{B} \\
\hline
\end{tabular}
\end{table}

The {\sc iraf} software suite \citep{tod86} was used to remove the
bias and pixel-to-pixel gain variations from each frame. Cosmic ray
events were removed using the technique described by \citet{cro95}, as
implemented in {\sc figaro} \citep{sho93}. The spectra were
straightened in {\sc figaro} so that the dispersion ran exactly along
rows of the image, then a two-dimensional wavelength fit was performed
to the arc images by fitting a third-order polynomial to the arc
wavelengths as a function of pixel number, for each row of the
image. These wavelength solutions were copied to the object images,
interpolating between the bracketing arc exposures, and the data were
rebinned so the wavelength-pixel relation was linear and uniform
across the image. The sky background was subtracted from each image,
by making a linear fit to the sky on either side of the galaxy,
choosing regions well outside the wings of the galaxy profile. The
spectra were corrected for atmospheric extinction, the telluric
absorption features were removed by comparing with the spectrum of a
smooth-spectrum standard taken at similar airmass, and the spectra
were flux-calibrated. We constructed two grand-sum spectra of the
nucleus -- one for each set-up (Table~\ref{tab:obs}) -- by aligning
the individual spectra and summing, then correcting the summed
spectrum for the Galactic reddening $A_V = 0.331\,\mathrm{mag}$\
\citep{sfd98}.

\begin{figure*}
     \centerline{\psfig{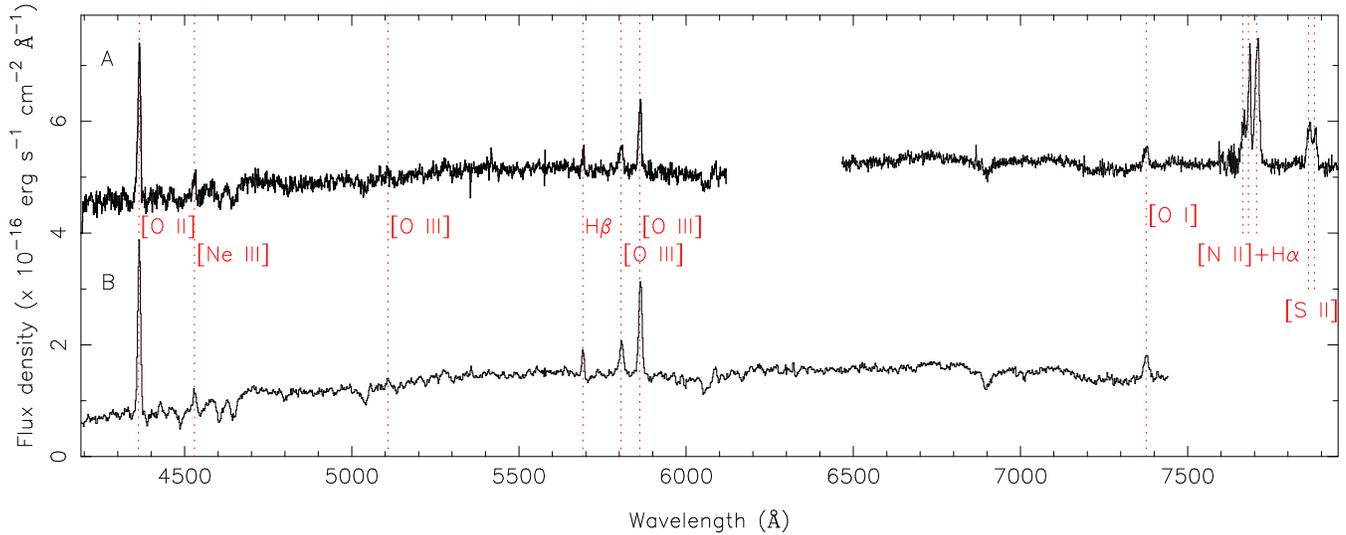}}
     \caption{Spectrum of the nucleus of MRC~B1221$-$423, extracted
       from the central 2\farcs7, with emission features identified.
       The top spectrum (spectrum A) is from the 2004 May 14 blue and
       red data, with the slit along the rotation axis of the galaxy;
       the bottom spectrum (spectrum B) is extracted from the combined
       data from 2004 May 15--16, with the slit along the line joining
       the galaxy to its companion.  Spectrum A is offset in the
       \textit{y}-direction by $4\times 10^{-16}\,
       \mathrm{erg}\,\mathrm{s}^{-1}\,\mathrm{cm}^{-2}\,\mathrm{\AA}^{-1}$\ 
       for clarity. }
     \label{fig:spectrum}
\end{figure*}

These spectra are shown in Figure~\ref{fig:spectrum}. They show narrow
emission lines from ionised gas, on top of a continuum dominated by
light from stars. In the following sections, we analyse these two
components separately.

\subsection{Stellar continuum}
\label{sec:Stellar-continuum}

\subsubsection{D4000 strength}
\label{sec:D4000}

\begin{figure}
     \centerline{\psfig{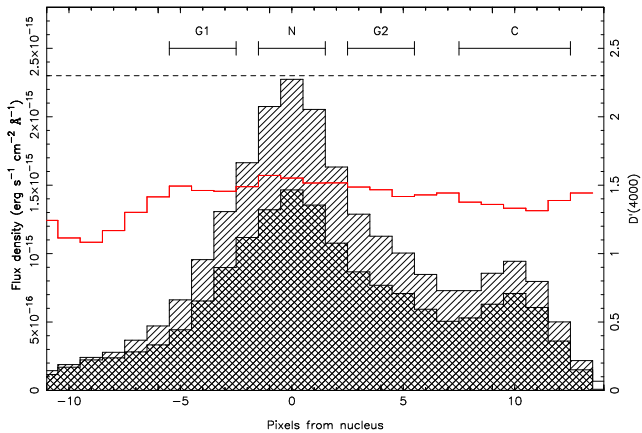}}
     \caption{Spatial profiles of the flux along the slit in spectrum
       A in the rest wavelength ranges 3750--3850$\,$\AA\
       (cross-hatching) and 4150--4250$\,$\AA\ (hatching). The ratio
       of the two, $D(4000)$, is shown as the solid line (right-hand
       scale); the dashed horizontal line shows $D(4000)=2.3$, which
       is the value expected for a 12.5~Gyr elliptical galaxy of solar
       metallicity. Sub-solar metallicity ($Z=0.4\;\mathrm{Z}_{\sun}$)
       lowers this ratio to $D(4000)=1.8$. The ratio measured for
       MRC~B1221$-$423 is consistently below this, indicating the
       presence of a blue spectral component. The horizontal bars at
       the top of the plot indicated the regions extracted for the
       nuclear spectrum N, the companion galaxy C, and two regions
       away from the nucleus G1 and G2.}
     \label{fig:D4000}
\end{figure}

The continua of many radio galaxies show UV excesses compared with
normal elliptical galaxies. This excess may arise either from
scattered nuclear light or from a population of young stars. In
Figure~\ref{fig:D4000}, we show the UV excess in MRC~B1221$-$423, as
measured by the strength of the 4000$\,$\AA\ break. We use $D'(4000)$\
as defined by \citet{tdm+02}, which is the ratio of the flux between
4150--4250$\,$\AA\ (above the 4000$\,$\AA\ break) to the flux between
3750--3850$\,$\AA\ (below the break). This is a version of the
original index $D(4000)$ proposed by \citet{bru83}, modified to
exclude emission lines which are often present in the spectra of radio
galaxies. Old stellar populations have large values of this index,
while smaller values indicate a less pronounced 4000$\,$\AA\ break,
and hence the presence of more UV light than expected from a passively
evolving stellar population.

Figure~\ref{fig:D4000} shows that there is an excess of UV flux in all
regions of MRC~B1221$-$423 compared to a typical elliptical galaxy.
Nowhere does the observed $D'(4000)$ approach the value of 2.3 typical
for a passively-evolving galaxy of solar metallicity \citep{trg+05};
the mean value over the galaxy (from the regions labelled G1 to G2) is
$1.48 \pm 0.05$.  \citet{tdm+02} measured $D'(4000)$ for a sample of
radio galaxies, and found values in the range $\sim 1.2$--2.6.
\citet{trg+05} found clear UV excesses in a sample of three powerful
radio galaxies, and interpreted this as evidence of star formation
associated with the radio source. The fact that we see a similar UV
excess throughout MRC~B1221$-$423 suggests that the star formation
associated with the radio source is galaxy-wide.

\subsubsection{Modelling the continuum}
\label{sec:Modelling-continuum}

Next, we model the shape of the continuum spectral energy distribution
in detail.  The spectrum from 2004 May 14 (spectrum A) had different
resolution, wavelength coverage, and position angle to the grand-sum
spectrum from the other two nights (spectrum B; Table~\ref{tab:obs}),
so we performed the following analysis for both summed spectra
separately.  We fitted three separate models to the continuum:

\begin{enumerate}
  \item a single-age stellar population, allowing the age to vary
  \item an old stellar population plus a power-law  continuum
  representing the AGN, with $F_\lambda \propto \lambda^\alpha$,
  allowing $\alpha$ to vary
  \item an old stellar population plus a single-age young stellar
  population, allowing the age of the young stars to vary.
\end{enumerate}

We used published isochrone spectral synthesis models from the
\textsc{gissel96} library \citep{bc93} for the stellar populations.
Both the ``old'' and ``young'' components were single-age stellar
populations, representing instantaneous bursts of star formation. We
used a Salpeter initial mass function with mass limits of 0.1 and
100\Msolar\ and solar metallicity, using the \citet{gs83} stellar
spectral atlas.  For the old population, we used an instantaneous
burst model with an age of 15~Gyr; the exact age of the ``old''
population is not critical, since the spectrum changes little at these
ages.  For the young stellar population, we used instantaneous burst
models with ages 10, 15, 20, 50, 80, 100, 150, \dots 5000~Myr, giving
us 14 models in total. We assembled the model spectra from the
\textsc{gissel96} library, and binned them to the same resolution as
the observed spectra.

\begin{table*}
\caption{Fits to the stellar continuum, in the nucleus and other
  regions of the galaxy. Fits are shown to spectrum A, taken along the
  rotation axis of the galaxy, and spectrum B, along the line joining
  the galaxy with its companion (lower resolution;
  Table~\protect{\ref{tab:obs}}). The three different models fitted to
  each spectrum are shown in column 2; column 3 shows the percentage
  of light at 4156$\,$\AA\ contributed by the blue component (where
  present); column 4 shows either the index of the power-law component
  or the age of the single-age stellar population component; column 5
  shows the reduced $\chi^2$\ of the best fit. Ranges on the
  parameters indicate 1--$\sigma$ errors on the best fit. The lower
  half of the table show the results of fits to the companion galaxy
  and to the main galaxy outside the nucleus. }
\label{tab:fits}
\begin{tabular}{llclc}
\hline
Spectrum   & Model                & \% blue component & Parameter &
$\chi^2_\nu$ \\
\hline
Nucleus (A)& Single age           & -           & 2.2--2.7 Gyr     & 4.18 \\
           & 15 Gyr + power-law   & $44 \pm 5$  & $\alpha=-0.54 \pm 0.55$  & 2.14 \\
           & 15 Gyr + young stars & $42 \pm 1$  & $< 23$ Myr & 1.35 \\
Nucleus (B)& Single age           & -  & 1.6--3 Gyr       & 2.10 \\
           & 15 Gyr + power-law   & $43 \pm 5$ & $\alpha=0.14 \pm 0.35$  & 3.51 \\
           & 15 Gyr + young stars & $43 \pm 1$ & 75--105 Myr & 1.63 \\
\\
Companion (B)   & 15 Gyr + young stars & $81 \pm 3$ & 150--200 Myr & 0.86 \\
Main galaxy (A) & \qquad '' \qquad ''  & $64 \pm 4$ & $< 110$ Myr  & 0.79 \\
Main galaxy (B) & \qquad '' \qquad ''  & $64 \pm 4$ & 140--180 Myr  & 1.10 \\
\hline
\end{tabular}
\end{table*}

The spectra (observed and model) were binned into 20$\,$\AA\ bins,
after which we removed bins corresponding to emission lines, residual
night-sky lines, cosmic rays and atmospheric absorption bands.  This
resulted in 83 wavelength bins for spectrum A and 113 bins for
spectrum B.  We measure the flux in these bins for both our observed
spectrum and models. The models were matched to the observed flux in a
bin at (rest-frame) 4165$\,$\AA; the normalisation was allowed to vary
between 75 per cent and 125 per cent of the flux of the observed
spectrum, to allow for possible differences in slope between the model
and the observed spectrum.  Varying the fraction of the blue
component, the normalisation, and the power-law index or age of the
young stars, the best-fitting model was found by calculating the
$\chi^2$ difference between the observed spectrum and each model
spectrum. We assumed errors of $\pm 4$~per cent in each wavelength
bin for the $\chi^2$-fitting.

\begin{figure*}
     \centerline{\psfig{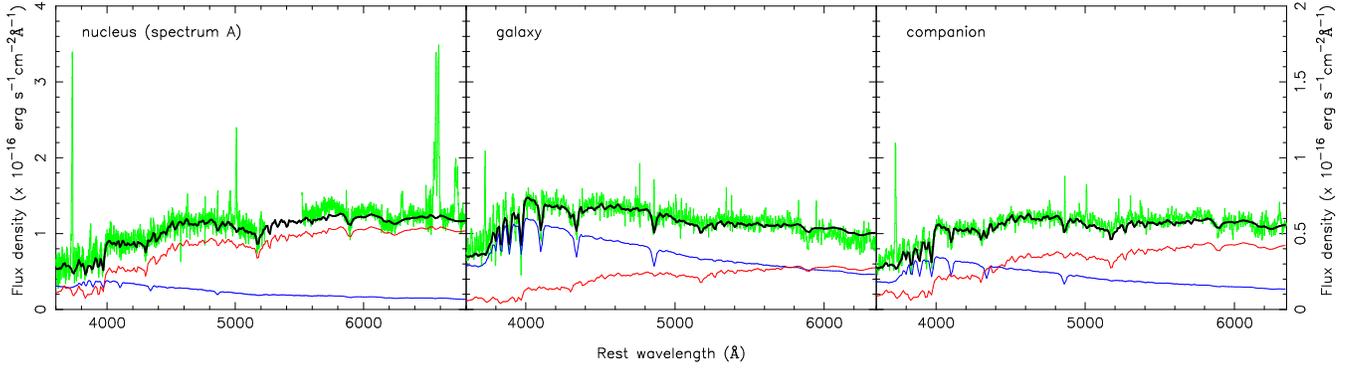}}
     \caption{The best two-component model fits to the continuum of
       the nucleus, the galaxy outside the nucleus (G1, north of the
       nucleus), and the companion galaxy; the emission lines were not
       included in the fits. The apertures used to extract the spectra
       are shown in Figure~\ref{fig:D4000}. Note that the $y$-scale of
       the left panel is double that of the other two panels
       (right-hand scale). The parameters of the fits are shown in
       \protect{Table~\ref{tab:fits}}. The model spectrum (thick black
       line) is shown on top of the observed spectrum, together with
       the individual components, the old population (red) and the
       young population (blue).}
     \label{fig:models}
\end{figure*}

We began by fitting to the spectrum of the nucleus of the galaxy,
extracting the central three rows (region N in Figure~\ref{fig:D4000}),
which correspond to the central 2.7~arcsec, or 7.8~kpc. The rest-frame
wavelength range over which the fitting was done was
3600--6100$\,$\AA\ for spectrum A, and 3300--6100$\,$\AA\ for spectrum
B. The results of this fitting are shown in Table~\ref{tab:fits}. No
single-age spectrum described the continuum well; the nominal best-fit
age of 2~Gyr systematically underestimates the flux in the red.
Spectrum A was reasonably well modelled by the addition of a power-law
with $\alpha=-1$, though the same model did not fit spectrum B well.
The addition of a population of young stars provided a much better fit
to the spectrum. Both spectra A and B are well fitted by a model where
$\sim 43$~per cent of the light at (rest-frame) 4200$\,$\AA\ is
contributed by a young stellar population with an age of $\la
100$~Myr.  Spectrum A requires that the young population have an age
of less than 23~Myr, while spectrum B was adequately fitted with
models having ages between 75 and 105~Myr. The fit to spectrum A is
shown in Figure~\ref{fig:models}.

Thus the continuum in the nuclear region is not well described by
either a single-age population, or a combination of old stars plus a
power-law component. Instead, we require at least two stellar
populations with different ages to reproduce the spectrum adequately.

We repeated this analysis with spectra extracted from other positions
along the slit. We extracted the spectrum at the position of the
companion galaxy; we also extracted regions of the main galaxy away
from the nucleus (Fig.~\ref{fig:models}). The spectrum of the
companion galaxy was extracted from spectrum B, including five rows
centred 9~arcsec south from the nucleus (region C in
Fig.~\ref{fig:D4000}).  For slit position B, we could extract spectra
from the southern and northern portions of the main galaxy separately.
We extracted three rows (2.7~arcsec) beginning $\pm 2.25$~arcsec
either side of the nucleus (regions G1 and G2 in Figure~\ref{fig:D4000})
and fitted the two spectra separately; the results for both northern
and southern spectra were consistent. For slit position A, the
signal-to-noise ratio of spectra extracted separately was
insufficient, so we extracted the same regions but added the spectra
together before fitting. As can be seen in Figure~\ref{fig:models},
the spectrum of the galaxy outside the nucleus shows a much younger
stellar population than the nucleus, with much stronger Balmer lines
and a smaller 4000$\,$\AA\ break.

To quantify this, we again fitted the observed spectra with two
single-age populations.  The results of this fitting are shown in
Table~\ref{tab:fits}.  The spectrum of the companion galaxy is
well-described by a model where 80 per cent of the light is contributed by a
200~Myr old stellar population.  The main galaxy away from the nucleus
is best fitted with a slightly younger second population, with ages
$\sim 100$~Myr. No significant difference was found between spectra
obtained on either side of the nucleus.  

We investigated including internal reddening as another free
parameter; the fits to the nuclear spectra were marginally improved by
the inclusion of 0.4 mag of extinction, but this did not significantly
change the fitted ages of the underlying populations. The fit to the
companion galaxy also indicates a reddening of $A_V \sim
0.4~\mathrm{mag}$, but the main galaxy outside the nucleus did not
demand any reddening.  These reddenings are significantly lower than
the values we derived in Paper I, where we modelled the pixel colours
using single-age populations.

This is broadly consistent with the results of Paper~I, based on
modelling of pixel colours, where we found an old population in the
`outskirts of the galaxy, an intermediate-age population ($\tau \sim
300$~Myr) in the companion and tidal tail, and a young population
($\tau \la 10$~Myr) in the nucleus and blue ``knots''. Modelling
the spectra enables us to best disentangle the contribution of the
different populations, although our spatial resolution is not as good
as in the images. Our fits to the pixel colours in the nucleus in
Paper~I were quite poor, indicating that our assumption of a
single-age population was not a good one. The two-age fit to the
spectrum in this work clearly gives a better description. 

Thus we  have found further evidence of  starburst activity throughout
the whole MRC~B1221$-$423 system, with stars of different ages present
in  different regions.  The youngest  stars are  in the  nucleus, with
slightly older populations in the companion galaxy and the main galaxy
outside of the nucleus. The stellar models enable us to determine what
fraction of  the mass of the  galaxy comes from  each population.  The
mass fraction in young stars is not large: the young population in the
nucleus with age $\la 75$~Myr, which contributes nearly half the light
at 4200$\,$\AA, contains less than 1 per cent of the mass. The highest
mass  fraction is  found in  the companion  galaxy, where  the 170~Myr
population contributing 80  per cent of the light makes  up 5 per cent
of the mass of the galaxy.

\subsubsection{Velocities}
\label{sec:Velocities}

We determined the rotation curve of the stellar population by
cross-correlating the spectrum extracted at different positions along
the slit with a synthetic spectrum from \citet{cbm+05}. We extracted
non-overlapping 2-pixel spectra along the slit in spectrum A, the
spectrum with the highest resolution, and cross-correlated with a
synthetic spectrum of an F5V star (with $T_\mathrm{eff}=6750$~K, $\log
g=4.5$), chosen as the best match to the continuum away from the
nucleus of the galaxy. We cross-correlated the region from rest-frame
3845--4800$\,$\AA, chosen to avoid the bright emission lines, using
the {\sc figaro} routine \texttt{scross}, which uses the method of
\citet{td79}. The resulting stellar rotation curve is shown as the
solid line in Figure~\ref{fig:vprofile}.

The redshift of the nucleus, which we take as the ``true'' redshift of
the galaxy, is $0.17053 \pm 0.00005$ ($cz = 51118 \pm 15$\kms). 

\subsection{Emission line gas}
\label{sec:Emission-line-gas}

The spectrum of MRC~B1221$-$423 shows numerous strong emission lines,
characteristic of powerful radio sources.  The emission lines enable
us to probe the physical conditions and kinematics of the gas in the
galaxy. Previous studies of CSS sources have found widespread evidence
of outflows and disturbed kinematics in these systems \citep[see
e.g.][]{htm03,emt+06}.

We subtracted the best fit to the continuum from each of Spectrum A
and spectrum B in order to model the emission lines. We used the fit
comprising two single-age populations, as determined in
\S~\ref{sec:Modelling-continuum} and Table~\ref{tab:fits}.

Table~\ref{tab:lineparams} shows the measured parameters of the
emission lines. The line centre and width (FWHM) were determined by
fitting a Gaussian to the subtracted spectrum; the line flux was
obtained by direct summation of the flux over a wavelength range four
times the FWHM, except where such a width would also include
neighbouring lines.

\begin{table}
\caption{Emission line measurements, from the rest-frame spectrum B
with the model stellar continuum subtracted.}
\label{tab:lineparams}
\begin{tabular}{l r@{$\;\pm\;$}l r@{$\;\pm\;$}l}
\hline
     & \multicolumn{2}{c}{Flux} & \multicolumn{2}{c}{FWHM} \\
Line & \multicolumn{2}{c}{($\times 10^{-16}\;\mathrm{erg}\;\mathrm{s}^{-1}\;\mathrm{cm}^{-2}$)} &
\multicolumn{2}{c}{(km s$^{-1}$)} \\
\hline
\oii\ 3727    & 37.0 & 1.5 & 660 & 40 \\
\neiii\ 3869  &  5.0 & 0.6 & 730 & 70 \\
\heii\ 4686   & \multicolumn{2}{c}{$< 0.12$} & \multicolumn{2}{c}{--} \\
H$\beta$ 4861 &  6.1 & 0.4 & 470 & 40 \\
\oiii\ 4959   &  8.6 & 0.2 & 600 & 30 \\
\oiii\ 5007   & 22.2 & 0.2 & 600 & 30 \\
\oi\ 6300     &  7.0 & 0.4 & 680 & 30 \\
\hline
\end{tabular}
\end{table}

A combined fit to the emission lines of the blue Spectrum A gives a
heliocentric redshift of $z=0.17068 \pm 0.00010$ ($cz = 51168 \pm
30$\kms); the red spectrum gives $z=0.17087 \pm 0.00013$ ($cz = 51225
\pm 40$\kms). The emission lines in Spectrum B give a redshift
consistent with those in blue Spectrum A: $z=0.17067 \pm 0.00010$
($cz=51166 \pm 30$\kms).

\subsubsection{Kinematics of the emission lines}
\label{sec:Kinem-emission-lines}

\begin{figure}
     \centerline{\psfig{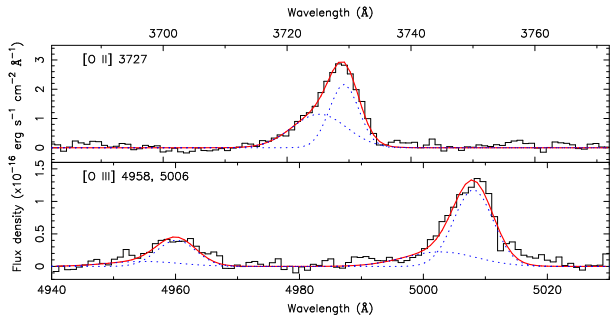}}
     \caption{Line profile fitted to the \oii\ and \oiii\ lines
     extracted from the central two rows of spectrum A. The dotted
     lines show the two separate Gaussian components which were
     fitted. The line separations and strengths of the \oiii\ doublet
     were constrained to those known from atomic physics, and we
     required the separation of the broad and narrow components to be
     the same as for the \oii\ line.}  \label{fig:emlines}
\end{figure}

\begin{figure}
     \centerline{\psfig{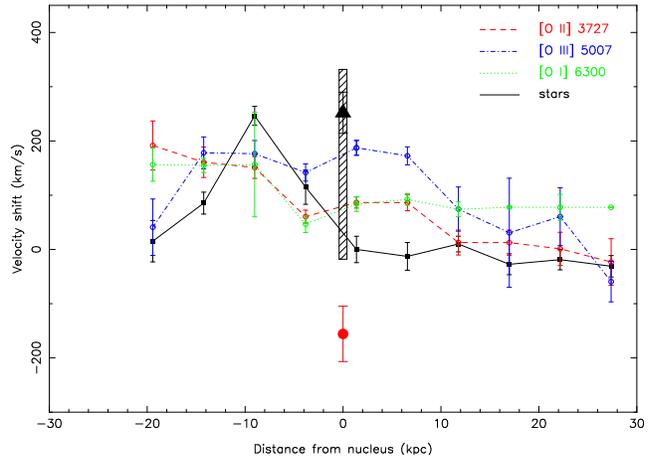}}
     \caption{Rest-frame velocity profiles along spectrum A, for the
       stellar continuum (solid line) and three emission lines:
       \oii~3727, \oiii~4959, 5007 and \oi~6300. The positive
       direction is to the south-east. Velocity zero is taken as the
       velocity of the absorption-line component at the position of
       the nucleus (corrected to the heliocentric frame, see
       \S~\protect{\ref{sec:Velocities}}). The large filled triangle
       shows the velocity of the narrow \hi\ absorption component
       (\S~\protect{\ref{sec:Radio-spectroscopy}}), while the large
       filled circle shows the velocity of the broad \oii\
       emission. The hatched bar shows the full width (FWZI) of the \hi\
       absorption. }
     \label{fig:vprofile}
\end{figure}

CSS sources often show broad, asymmetric emission lines in the nuclear
region. We extracted the spectrum at the position of the nucleus by
selecting the central two rows from each two-dimensional spectrum,
corresponding to a linear size of 5.2$\;$kpc. The brightest line,
\oii\,3727, has a very asymmetric blue wing, as shown in
Figure~\ref{fig:emlines}. We modelled the line with two gaussians, a
narrow component (with width $\Delta v = 450$\kms\ FWHM) at a velocity
very similar to the rest-frame velocity, and a broad component
($\Delta v = 830$\kms\ FWHM) blueshifted by 170\kms. The velocity of
this component is shown as the large red circle in
Figure~\ref{fig:vprofile}.

The \oiii\ 4959, 5007 doublet also showed an asymmetric blue wing in
the nuclear spectrum, though the broad component was noticeably weaker,
compared to the narrow component, than in the \oii\ 3727 line. We fitted
two Gaussian components to each line, constraining the wavelength
ratio to the known value and the flux ratios to be 3.0:1, given by the
transition probabilities; we also constrained the corresponding
components (broad and narrow) to have the same line width in each line
of the doublet. The widths of the gaussians were consistent with the
line widths from the fit to the \oii\ line, so we constrained them to
be the same, allowing only the line intensities to vary. The resulting
fit was also acceptable, though with a much weaker broad
component. None of the other emission lines --- H$\beta$, \neiii, \oi
--- showed any evidence for the presence of a broad component, so if
present it must be extremely weak.

To investigate the gas kinematics away from the nucleus, we extracted
spectra from different positions along the slit, as described in
\S~\ref{sec:Velocities}. Except for the nuclear spectrum, there was no
evidence for a broad component, so we fitted a single gaussian to each
line. The resulting velocity profiles are shown in
Figure~\ref{fig:vprofile}.

\subsubsection{Ionisation state}

The ionisation state of the MRC~B1221$-$423 is quite low; \oii~3727 is
stronger than \oiii~5007, while \heii~4686 is extremely weak. The
line-ratios are characteristic of ``low ionisation nuclear
emission-line regions'' or LINERs \citep{hec80}. The
temperature-sensitive \oiii\ line intensity ratio ($\lambda$
5007+$\lambda$ 4959)/$\lambda$ 4363 is relatively low, indicative of
high temperatures and/or densities.


The H$\alpha$\ to H$\beta$ ratio is extreme; we measure
$F_\mathrm{H\alpha}/F_\mathrm{H\beta}=6.36$, much larger than the case
B recombination value of 2.86 \citep[e.g.][]{ost89}. If this were due
to reddening, it would imply $A_V = 2.3$~mag, which is much higher
than the value we derived from the stellar continuum, $A_V \sim
0.4$~mag (\S~\ref{sec:Modelling-continuum}). Interestingly, this is
similar to our earlier estimate of the reddening in the nucleus from
modelling the spectral energy distributions (Paper~I), where we found
$A_V = 1.9$~mag. If the Balmer emission is mainly from the nucleus, we
could be seeing the effect of very localised dust.  On the other hand,
this ratio could be being distorted by the presence of
H$\beta$\ in absorption; for comparison, we measure
$F_\mathrm{H\gamma}/F_\mathrm{H\beta}=0.48$, compared to an expected
value of 0.47 for $T=10^4~\mathrm{K}$. Large ratios of
H$\alpha$/H$\beta$ are seen in many LINER sources \citep{hec80}.

\begin{figure}
     \centerline{\psfig{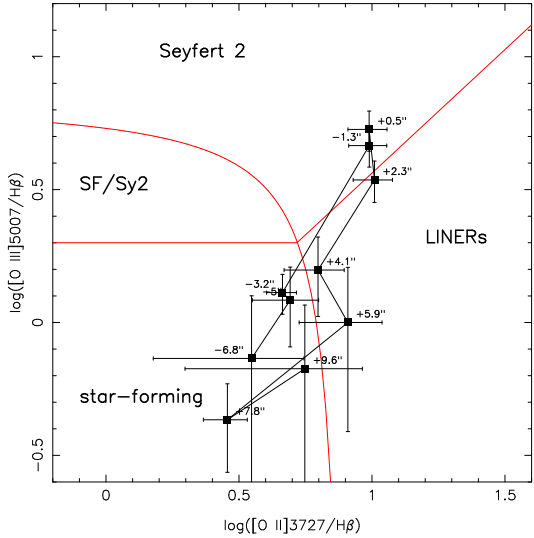}}
     \caption{Ionisation mechanism diagnostic plot using the line
       ratios \protect{\oii}3727/H$\beta$ and
       \protect{\oiii}5007/H$\beta$. Data points represent the
       observed spectra extracted in non-overlapping 2-pixel steps
       along the slit in spectrum A. The red lines show the
       classification regions from Lamareille (2010). The emission
       lines are AGN-like close to the core, and become more
       star-formation-like further out.}
     \label{fig:diag}\nocite{lam10}
\end{figure}

We can investigate the changing ionisation state of the gas by
comparing emission-line intensity ratios at different positions along
the slit \citep{bpt81}. In Figure~\ref{fig:diag} we show the
diagnostic diagram for the line ratios \oii3727/H$\beta$ and
\oiii5007/H$\beta$, using the gaussian fits to the lines described in
\S~\ref{sec:Kinem-emission-lines}. We also show the classification
regions of \citet{lam10}.  The diagram clearly shows the changing
ionisation state at different distances from the nucleus. Close to the
nucleus the line-ratios are consistent with ionisation by the AGN,
while at larger distances the line ratios look more like star-forming
galaxies.

\section{Discussion and Conclusions}
\label{sec:discussion}

We have presented detailed radio and optical spectroscopic
observations of MRC~B1221$-$423, a young CSS source which is currently
undergoing a merger. Our main findings are
\begin{itemize}
\item a highly distorted radio morphology
\item a large Faraday depth and significant depolarisation in the
northern, less distorted, lobe of the source
\item the presence of redshifted \hi\ in absorption
\item a range of stellar ages in the galaxy
\item strong emission lines with relatively low ionisation.
\end{itemize}

The radio observations of MRC~B1221$-$423 show dramatic interaction
between the southern radio jet and the host galaxy, together with
infalling gas. This jet appears to be bent through a full 180\degr. The
alignment of the magnetic field along the jet supports the
interpretation of the feature as a single connected structure, with
the magnetic field being oriented circumferentially where the extreme
bending occurs. This is suggestive of a strong interaction with high
density material. Interestingly, the RM synthesis shows no evidence of
a dense interstellar medium where the jet bends.

The much shorter extent of the radio emission to the north, as well as
the steep gradient in Faraday depth from north to south, suggests the
presence of a dense and possibly highly magnetised medium local to the
source. One possibility is that we could be seeing an orientation
effect, with the northern jet pointed away from us and seen behind the
galaxy. Large changes of jet direction in radio galaxies have been
modeled as the result of a helical jet \citep[see
e.g.][]{hmcp84,cm93}; it is possible that a helix viewed almost
head-on could explain the extreme change in projected jet direction
that we see in MRC~B1221$-$423. The lack of evidence for a dense ISM
at the bend of the jet from the RM synthesis would be consistent with
this interpretation.  Alternately, the infalling gas could be
interacting with the jet, producing both the change in jet direction
and the different polarisation in the lobes.  \citet{sg03} found a
high degree of polarisation asymmetry in CSS sources, which they
ascribed to interaction of the jets with infalling material.

\hi\ was clearly detected in absorption, with a narrow component at a
velocity $+250\kms$\ compared to the velocity of the galaxy determined
from the stellar component. Both this absorption and the radio
continuum observations suggest the source is in a gas-rich
environment. There is also a broader wing of \hi\ absorption extending
to the systemic velocity.  The velocity shift with respect to the
stellar redshift suggests large amounts of infalling gas. It should be
noted, however, that the velocities of the \hi, particularly of the
broad component, are quite similar to the range of velocities shown by
the emission lines (see Fig.~\ref{fig:vprofile}), so it is possible
the \hi\ absorption is just part of the overall distribution of gas.
Redshifted \hi\ absorption is not common in similar sources: of 14
sources studied by \citet{htm08}, most showed significantly
\textit{blueshifted} \hi\ absorption, reflecting outflow. Only two of
their sources --- PKS~B0023$-$263 and PKS~B1934$-$638 --- showed
redshifted \hi\ absorption.

The location of the absorbing gas is not clear. A comparison can be
drawn with the radio galaxy NGC~315, which shows two absorption
systems against the central region \citep{mpo+09}.  In addition to the
redshifted narrow component ($v \sim +490$\kms), NGC~315 has a broader
(FWZI $\sim 150$\kms) component redshifted by $\sim 80$\kms\ compared
to the systemic velocity. \citet{mpo+09} concluded from velocity
gradients seen in VLBI data that the broad component arises from gas
falling into the nucleus, while the narrow component arises from more
distant gas clouds falling towards the galaxy. By analogy, we might be
seeing a similar situation in MRC~B1221$-$423. The narrow absorption
component could represent a discrete cloud at some distance from the
galaxy, while the broad absorption arises much closer to the nucleus,
either from a circumnuclear torus, or from the infalling gas that is
fueling the nuclear activity. Higher resolution \hi\ observations are
needed to distinguish these possibilities.

Optical spectra of MRC~B1221$-$423 reveal disturbed kinematics. Most
of the ionised gas is at velocities similar to that of the stars, but
in the nucleus we detect a broad blue-shifted component to the
emission lines with $\Delta v = 830$\kms\ FWHM, which we interpret as
evidence for outflow of emission-line gas in the nucleus.  Since it is
only visible in the nucleus, this could arise from gas interacting
with the radio jet: the fact that it is blue-shifted might support the
interpretation that the jet axis is strongly inclined to the line of
sight.

The ionisation state of the gas is relatively low, with \oii~3727
stronger than \oiii~5007, and a very weak \heii~4686 line. The line
ratios change as a function of position, with the emission being
dominated by the AGN close to the nucleus, and becoming more
star-formation-like further out.  A similar result was found in the
sample of compact radio sources studied by \citet{htm09}, where the
ionisation in the nuclear regions is dominated by the central
AGN. Like them, we find a lower \oiii~5007 luminosity than extended
sources with similar radio luminosity in their comparison sample.

Star-formation activity is evident over the whole system, including
the companion galaxy. The youngest stellar population is close to the
nucleus, consistent with the idea that the same inflows of gas that
have triggered star formation are also triggering the activity of the
central black hole.

Our previous work on MRC~B1221$-$423 suggested the following scenario:
tidal interaction with the companion galaxy triggered different phases
of star formation in the two galaxies, and then, after a substantial
delay, the central AGN was triggered to produce the radio emission.
The current observations support this model. We have found evidence of
past episodes of star formation, probable evidence for infalling gas,
and indication of interaction between the radio jet and its
environment.

Our work confirms the fact that young, powerful radio sources tend to
live in a gas-rich nuclear environment. Further, we show that the AGN
is younger than the starburst.  The ages of the stellar populations
are comparable with the orbital period of the companion, which we
estimate to be $\sim 10^8$~yr, based on the projected separation of
the companion galaxy from the nucleus, together with an estimate of
the mass of the galaxy from the $K$-band luminosity. The faint arcs to
the north-east \citep{shp03} suggest there have been previous tidal
interactions. These timescales suggest a time delay of $\sim$~a few
$\times 100$~Myr between the tidal event (that triggered star
formation) and the onset of AGN activity.  Such delays have been found
in several nearby AGN \citep[see e.g.][]{tds96,emt+06,dsm+07}, but
this is possibly one of the best examples.  In recent work,
\citet{whc10} constructed a sample of starburst galaxies from the
Sloan Digital Sky Survey. They found that the rate of accretion onto
the central black hole, as measured by the strength of the \oiii\
line, rises steeply about 250~Myr after the onset of star formation.
They speculate that the accretion onto the black hole may be being
regulated by the starburst, with the accretion being dominated by
low-velocity stellar winds, and accretion at early times being
suppressed by supernovae.

We have also found indication of a possible infall of cold gas that
could provide the fuel for the AGN. We note that the age distribution
of the stellar populations is consistent with a picture in which cold
or cooling gas is driven by the merger towards the centre of the host
galaxy to fuel the AGN from relatively large radii, and that this gas
begins to form stars while it is still well away from the
nucleus. This contrasts with the situation in lower-luminosity radio
sources in cluster environments, which, even though they sometimes
contain the most massive black holes, are accreting in an environment
dominated by hot gas and are associated with an accretion mode that
produces low-power radio jets.

\section*{Acknowledgments}

We thank the anonymous referee for helpful comments.  The Australia
Telescope is funded by the Commonwealth of Australia for operation as
a National Facility managed by CSIRO.  RWH and HMJ acknowledge support
from the Australian Research Council.

This research has made use of the NASA/IPAC Extragalactic Database
(NED) which is operated by the Jet Propulsion Laboratory, California
Institute of Technology, under contract with the National Aeronautics
and Space Administration. It has also made use of the VizieR catalogue
access tool, CDS, Strasbourg, France, and Ned Wright's Javascript
Cosmology Calculator \citep{wri06}.


\begin{thebibliography}{}

\bibitem[\protect\citeauthoryear{Baldwin, Phillips \& Terlevich}{Baldwin
  et~al.}{1981}]{bpt81}
Baldwin J.~A.,  Phillips M.~M.,    Terlevich R.,  1981, PASP, 93, 5

\bibitem[\protect\citeauthoryear{Barnes \& Hernquist}{Barnes \&
  Hernquist}{1992}]{bh92}
Barnes J.~E.,  Hernquist L.,  1992, ARA\&A, 30, 705

\bibitem[\protect\citeauthoryear{Bolton \& Shimmins}{Bolton \&
  Shimmins}{1973}]{bs73}
Bolton J.~G.,  Shimmins A.~J.,  1973, Aust. J. Phys. Astr. Supp., 30, 1

\bibitem[\protect\citeauthoryear{Brentjens \& de Bruyn}{Brentjens \&
  de~Bruyn}{2005}]{bb05}
Brentjens M.~A.,  de Bruyn A.~G.,  2005, A\&A, 441, 1217

\bibitem[\protect\citeauthoryear{Bruzual}{Bruzual}{1983}]{bru83}
Bruzual G.,  1983, ApJ, 273, 105

\bibitem[\protect\citeauthoryear{Bruzual \& Charlot}{Bruzual \&
  Charlot}{1993}]{bc93}
Bruzual G.,  Charlot S.,  1993, ApJ, 405, 538

\bibitem[\protect\citeauthoryear{Burgess \& Hunstead}{Burgess \&
  Hunstead}{2006a}]{bh06}
Burgess A.~M.,  Hunstead R.~W.,  2006a, AJ, 131, 100

\bibitem[\protect\citeauthoryear{Burgess \& Hunstead}{Burgess \&
  Hunstead}{2006b}]{bh06b}
Burgess A.~M.,  Hunstead R.~W.,  2006b, AJ, 131, 114

\bibitem[\protect\citeauthoryear{Coelho, Barbuy, Mel{\'e}ndez, Schiavon \&
  Castilho}{Coelho et~al.}{2005}]{cbm+05}
Coelho P.,  Barbuy B.,  Mel{\'e}ndez J.,  Schiavon R.~P.,    Castilho B.~V.,
  2005, A\&A, 443, 735

\bibitem[\protect\citeauthoryear{Conway \& Murphy}{Conway \&
  Murphy}{1993}]{cm93}
Conway J.~E.,  Murphy D.~W.,  1993, ApJ, 411, 89

\bibitem[\protect\citeauthoryear{Croke}{Croke}{1995}]{cro95}
Croke B. F.~W.,  1995, PASP, 107, 1255

\bibitem[\protect\citeauthoryear{Croton, Springel, White, De~Lucia, Frenk, Gao,
  Jenkins, Kauffmann, Navarro \& Yoshida}{Croton et~al.}{2006}]{csw+06}
Croton D.~J.,  Springel V.,  White S. D.~M.,  De~Lucia G.,  Frenk C.~S.,  Gao
  L.,  Jenkins A.,  Kauffmann G.,  Navarro J.~F.,    Yoshida N.,  2006, MNRAS,
  365, 11

\bibitem[\protect\citeauthoryear{Davies, S{\'a}nchez, Genzel, Tacconi, Hicks,
  Friedrich \& Sternberg}{Davies et~al.}{2007}]{dsm+07}
Davies R.~I.,  S{\'a}nchez F.~M.,  Genzel R.,  Tacconi L.~J.,  Hicks E. K.~S.,
  Friedrich S.,    Sternberg A.,  2007, ApJ, 671, 1388

\bibitem[\protect\citeauthoryear{Duncan \& Sproats}{Duncan \&
  Sproats}{1992}]{ds92}
Duncan R.~A.,  Sproats L.~N.,  1992, PASA, 10, 16

\bibitem[\protect\citeauthoryear{Emonts, Morganti, Tadhunter, Holt, Oosterloo,
  van~der Hulst \& Wills}{Emonts et~al.}{2006}]{emt+06}
Emonts B. H.~C.,  Morganti R.,  Tadhunter C.~N.,  Holt J.,  Oosterloo T.~A.,
  van~der Hulst J.~M.,    Wills K.~A.,  2006, A\&A, 454, 125

\bibitem[\protect\citeauthoryear{Gooch}{Gooch}{1996}]{goo96}
Gooch R.,  1996, in Jacoby G.~H.,  Barnes J.,  eds, Astronomical Data Analysis
  Software and Systems V, ASP Conf. Series, 101, 80

\bibitem[\protect\citeauthoryear{Gregory, Vavasour, Scott \& Condon}{Gregory
  et~al.}{1994}]{gvsc94}
Gregory P.~C.,  Vavasour J.~D.,  Scott W.~K.,    Condon J.~J.,  1994, ApJS, 90,
  173

\bibitem[\protect\citeauthoryear{Gunn \& Stryker}{Gunn \& Stryker}{1983}]{gs83}
Gunn J.~E.,  Stryker L.~L.,  1983, ApJS, 52, 121

\bibitem[\protect\citeauthoryear{Heald, Braun \& Edmonds}{Heald
  et~al.}{2009}]{hbe09}
Heald G.,  Braun R.,    Edmonds R.,  2009, A\&A, 503, 409

\bibitem[\protect\citeauthoryear{Heckman}{Heckman}{1980}]{hec80}
Heckman T.~M.,  1980, A\&A, 87, 152

\bibitem[\protect\citeauthoryear{Heckman, Smith, Baum, van Breugel, Miley,
  Illingworth, Bothun \& Balick}{Heckman et~al.}{1986}]{hsb+86}
Heckman T.~M.,  Smith E.~P.,  Baum S.~A.,  van Breugel W. J.~M.,  Miley G.~K.,
  Illingworth G.~D.,  Bothun G.~D.,    Balick B.,  1986, ApJ, 311, 526

\bibitem[\protect\citeauthoryear{Holt}{Holt}{2009}]{hol09}
Holt J.,  2009, AN, 330, 226

\bibitem[\protect\citeauthoryear{Holt, Tadhunter \& Morganti}{Holt
  et~al.}{2003}]{htm03}
Holt J.,  Tadhunter C.~N.,    Morganti R.,  2003, MNRAS, 342, 227

\bibitem[\protect\citeauthoryear{Holt, Tadhunter \& Morganti}{Holt
  et~al.}{2008}]{htm08}
Holt J.,  Tadhunter C.~N.,    Morganti R.,  2008, MNRAS, 387, 639

\bibitem[\protect\citeauthoryear{Holt, Tadhunter \& Morganti}{Holt
  et~al.}{2009}]{htm09}
Holt J.,  Tadhunter C.~N.,    Morganti R.,  2009, MNRAS, 400, 589

\bibitem[\protect\citeauthoryear{Hunstead, Murdoch, Condon \&
  Phillips}{Hunstead et~al.}{1984}]{hmcp84}
Hunstead R.~W.,  Murdoch H.~S.,  Condon J.~J.,    Phillips M.~M.,  1984, MNRAS,
  207, 55

\bibitem[\protect\citeauthoryear{Johnston, Hunstead, Cotter \& Sadler}{Johnston
  et~al.}{2005}]{jhcs05}
Johnston H.~M.,  Hunstead R.~W.,  Cotter G.,    Sadler E.~M.,  2005, MNRAS,
  356, 515

\bibitem[\protect\citeauthoryear{Lamareille}{Lamareille}{2010}]{lam10}
Lamareille F.,  2010, A\&A, 509, A53

\bibitem[\protect\citeauthoryear{Large, Mills, Little, Crawford \&
  Sutton}{Large et~al.}{1981}]{lml+81}
Large M.~I.,  Mills B.~Y.,  Little A.~G.,  Crawford D.~F.,    Sutton J.~M.,
  1981, MNRAS, 194, 693

\bibitem[\protect\citeauthoryear{Mauch, Murphy, Buttery, Curran, Hunstead,
  Piestrzynski, Robertson \& Sadler}{Mauch et~al.}{2003}]{mmb+03}
Mauch T.,  Murphy T.,  Buttery H.~J.,  Curran J.,  Hunstead R.~W.,
  Piestrzynski B.,  Robertson J.~G.,    Sadler E.~M.,  2003, MNRAS, 342, 1117

\bibitem[\protect\citeauthoryear{Morganti, Peck, Oosterloo, van Moorsel,
  Capetti, Fanti, Parma \& de Ruiter}{Morganti et~al.}{2009}]{mpo+09}
Morganti R.,  Peck A.~B.,  Oosterloo T.~A.,  van Moorsel G.,  Capetti A.,
  Fanti R.,  Parma P.,    de Ruiter H.~R.,  2009, A\&A, 505, 559

\bibitem[\protect\citeauthoryear{Morganti, Tadhunter \& Oosterloo}{Morganti
  et~al.}{2005}]{mto05}
Morganti R.,  Tadhunter C.~N.,    Oosterloo T.~A.,  2005, A\&A, 444, L9

\bibitem[\protect\citeauthoryear{Murphy, Mauch, Green, Hunstead, Piestrzynska,
  Kels \& Sztajer}{Murphy et~al.}{2007}]{mmg+07}
Murphy T.,  Mauch T.,  Green A.,  Hunstead R.~W.,  Piestrzynska B.,  Kels
  A.~P.,    Sztajer P.,  2007, MNRAS, 382, 382

\bibitem[\protect\citeauthoryear{Murphy et al.}{Murphy et
    al.}{2010}]{mse+10} 
Murphy T.,  Sadler E.~M.,  Ekers R.~D.,  Massardi M.,  et al. 2010,
MNRAS, 402, 2403

\bibitem[\protect\citeauthoryear{O'Dea}{O'Dea}{1998}]{ode98}
O'Dea C.~P.,  1998, PASP, 110, 493

\bibitem[\protect\citeauthoryear{Osterbrock}{Osterbrock}{1989}]{ost89}
Osterbrock D.~E.,  1989, Astrophysics of gaseous nebulae and active galactic
  nuclei.
University Science Books, Mill Valley

\bibitem[\protect\citeauthoryear{Rawlings \& Saunders}{Rawlings \&
  Saunders}{1991}]{rs91}
Rawlings S.,  Saunders R.,  1991, Nat, 349, 138

\bibitem[\protect\citeauthoryear{Ricci, Prandoni, Gruppioni, Sault \& de
  Zotti}{Ricci et~al.}{2006}]{rpg+06}
Ricci R.,  Prandoni I.,  Gruppioni C.,  Sault R.~J.,    de Zotti G.,  2006,
  A\&A, 445, 465

\bibitem[\protect\citeauthoryear{Safouris, Hunstead \& Prouton}{Safouris
  et~al.}{2003}]{shp03}
Safouris V.,  Hunstead R.~W.,    Prouton O.~R.,  2003, PASA, 20, 1

\bibitem[\protect\citeauthoryear{Saikia \& Gupta}{Saikia \& Gupta}{2003}]{sg03}
Saikia D.~J.,  Gupta N.,  2003, A\&A, 405, 499

\bibitem[\protect\citeauthoryear{Sault, Teuben \& Wright}{Sault
  et~al.}{1995}]{stw95}
Sault R.~J.,  Teuben P.~J.,    Wright M. C.~H.,  1995, in Shaw R.~A.,  Payne
  H.~E.,   Hayes J. J.~E.,  eds, Astronomical Data Analysis Software and
  Systems IV, ASP Conf. Series, 77, 433

\bibitem[\protect\citeauthoryear{Schlegel, Finkbeiner \& Davis}{Schlegel
  et~al.}{1998}]{sfd98}
Schlegel D.~J.,  Finkbeiner D.~P.,    Davis M.,  1998, ApJ, 500, 525

\bibitem[\protect\citeauthoryear{Shortridge}{Shortridge}{1993}]{sho93}
Shortridge K.,  1993, in Hanisch R.~J.,  Brissenden R. J.~V.,   Barnes J.,
  eds, Astronomical Data Analysis Software and Systems II, ASP
  Conf. Series, 52, 219

\bibitem[\protect\citeauthoryear{Simpson, Clements, Rawlings \& Ward}{Simpson
  et~al.}{1993}]{scrw93}
Simpson C.,  Clements D.~L.,  Rawlings S.,    Ward M.,  1993, MNRAS, 262, 889

\bibitem[\protect\citeauthoryear{Slee}{Slee}{1977}]{sle77}
Slee O.~B.,  1977, Aust. J. Phys. Astr. Supp., 43, 1

\bibitem[\protect\citeauthoryear{Slee}{Slee}{1995}]{sle95}
Slee O.~B.,  1995, Aust. J. Phys., 48, 143

\bibitem[\protect\citeauthoryear{Slee \& Higgins}{Slee \&
  Higgins}{1973}]{sh73b}
Slee O.~B.,  Higgins C.~S.,  1973, Aust. J. Phys. Astr. Supp., 27, 1

\bibitem[\protect\citeauthoryear{Slee \& Siegman}{Slee \& Siegman}{1988}]{ss88}
Slee O.~B.,  Siegman B.~C.,  1988, MNRAS, 235, 1313

\bibitem[\protect\citeauthoryear{Tadhunter, Dickson, Morganti, Robinson, Wills,
  Villar-Mart{\'\i}n \& Hughes}{Tadhunter et~al.}{2002}]{tdm+02}
Tadhunter C.,  Dickson R.,  Morganti R.,  Robinson T.~G.,  Wills K.,
  Villar-Mart{\'\i}n M.,    Hughes M.,  2002, MNRAS, 330, 977

\bibitem[\protect\citeauthoryear{Tadhunter, Robinson, Gonz{\'a}lez~Delgado,
  Wills \& Morganti}{Tadhunter et~al.}{2005}]{trg+05}
Tadhunter C.,  Robinson T.~G.,  Gonz{\'a}lez~Delgado R.~M.,  Wills K.,
  Morganti R.,  2005, MNRAS, 356, 480

\bibitem[\protect\citeauthoryear{Tadhunter, Dickson \& Shaw}{Tadhunter
  et~al.}{1996}]{tds96}
Tadhunter C.~N.,  Dickson R.~C.,    Shaw M.~A.,  1996, MNRAS, 281, 591

\bibitem[\protect\citeauthoryear{Tody}{Tody}{1986}]{tod86}
Tody D.,  1986, in Crawford D.,  ed., Instrumentation in Astronomy VI Vol.~627,
  pp 733--752

\bibitem[\protect\citeauthoryear{Tonry \& Davis}{Tonry \& Davis}{1979}]{td79}
Tonry J.,  Davis M.,  1979, AJ, 84, 1511

\bibitem[\protect\citeauthoryear{Wild, Heckman \& Charlot}{Wild
  et~al.}{2010}]{whc10}
Wild V.,  Heckman T.,    Charlot S., 2010, Timing the starburst-{AGN}
  connection, MNRAS, in press (arXiv:1002.3156)

\bibitem[\protect\citeauthoryear{Wills}{Wills}{1975}]{wil75}
Wills B.~J.,  1975, Aust. J. Phys. Astr. Supp., 38, 1

\bibitem[\protect\citeauthoryear{Wright, Griffith, Burke \& Ekers}{Wright
  et~al.}{1994}]{wgbe94}
Wright A.~E.,  Griffith M.~R.,  Burke B.~F.,    Ekers R.~D.,  1994, ApJS, 91,
  111

\bibitem[\protect\citeauthoryear{Wright \& Otrupcek}{Wright \&
  Otrupcek}{1990}]{wo90}
Wright A.,  Otrupcek R., 1990, {P}arkes Catalog 1990, Australia Telescope
  National Facility

\bibitem[\protect\citeauthoryear{Wright, Wark, Troup, Otrupcek, Hunt \&
  Cooke}{Wright et~al.}{1990}]{wwt+90}
Wright A.~E.,  Wark R.~M.,  Troup E.,  Otrupcek R.,  Hunt A.,    Cooke D.~J.,
  1990, PASA, 8, 261

\bibitem[\protect\citeauthoryear{Wright}{Wright}{2006}]{wri06}
Wright E.~L.,  2006, PASP, 118, 1711

\end{thebibliography}

\appendix \section{Flux measurements} In
Table~\ref{table:radio_fluxes} we present radio flux measurements of
B1221$-$423 from the literature.

\begin{table*}
\begin{minipage}{12cm}
\caption{B1221$-$423 radio flux densities.}
\label{table:radio_fluxes}
\begin{tabular}{@{}rrcl}
\hline
\multicolumn{1}{c}{Frequency} & \multicolumn{1}{c}{$S_{\nu}$} & Survey/ & Reference \\
 \multicolumn{1}{c}{(MHz)} & \multicolumn{1}{c}{(mJy)} & Telescope &  \\
\hline
80 &  $8000 \pm	1300$\rlap{$^a$} & CCA & \citet{sh73b,sle95} \\
160 & $8100 \pm	800$\rlap{$^a$} & CCA & \citet{sle77,sle95} \\
408 & $5080 \pm	130$ & MRC & \citet{lml+81} \\
468 & 	$4670 \pm	810$ & PKS & \citet{wil75} \\
635 & 	$3910 \pm	270$ & PKS & \citet{wil75} \\
843 & 	$3300 \pm	200$ & MOST & \citet{bh06} \\
843 & 	$3110 \pm	90$ & SUMSS & \citet{mmb+03,mmg+07} \\
960 & 	$3070 \pm	170$ & PKS & \citet{wil75} \\
1344 & 	$2360 \pm	50$ & ATCA & This paper \\
1382 & 	$2290 \pm	70$ & ATCA & This paper \\
1410 & 	$2470 \pm	130$ & PKS & \citet{wil75} \\
2280 & 	$1650 \pm	80$ & ATCA & This paper \\
2496 & 	$1600 \pm	50$ & ATCA & This paper \\
2650 & 	$1700 \pm 	50$ & PKS & \citet{wil75} \\
2700 & 	$1620 \pm	50$ & PKS & \citet{bs73} \\
4740 & 	$1050 \pm	50$\rlap{$^b$} & ATCA & \citet{bh06b} \\
4790 & 	$1010 \pm	50$\rlap{$^{bc}$} & ATCA & \citet{ds92} \\
4800 & 	$1010 \pm	20$ & ATCA & \citet{shp03} \\
4800 &  $999  \pm       41$ & ATCA & \citet{mse+10} \\
4850 & 	$1090 \pm	60$\rlap{$^d$} & PMN & \citet{wgbe94} \\
4850 & 	$1140 \pm	60$ & PMNM & \citet{gvsc94} \\
5009 & 	$1000 \pm	50$ & PKS & \citet{wil75} \\
8400 & 	$630 \pm	70$ & PKS & \citet{wwt+90,wo90} \\
8640 & 	$610 \pm	20$ & ATCA & \citet{shp03} \\
18\,500 & 	$357 \pm	18$ & ATCA & \citet{rpg+06} \\
18\,880.5 & 	$310 \pm	30$ & ATCA & This paper \\
19\,648.5 & 	$290 \pm	30$ & ATCA & This paper \\
19\,904 &  $330  \pm       17$ & AT20G & \citet{mse+10} \\
22\,000 & 	 $347 \pm	35$ & ATCA & \citet{rpg+06} \\
\hline
\end{tabular}
\end{minipage}

\medskip
\begin{minipage}{12cm}
Notes: CCA -- Culgoora Circular Array; MRC 
  -- Molonglo Reference Catalogue; MOST -- Molonglo Observatory
  Synthesis Telescope; SUMSS -- Sydney University Molonglo Sky Survey;
  PKS -- Parkes Catalogue; ATCA -- Australia Telescope Compact Array;
  PMN -- Parkes-MIT-NRAO Survey; PMNM -- PMN Map Catalogue; AT20G --
  Australia Telescope 20 GHz Survey \hfill

$^a$Flux density uncertainty calculated using the error analysis from
the original CCA catalogue, but not including the original uncertainty
associated with the flux density scale; the scale was refined by
\citet{ss88}. \hfill

$^b$Flux density uncertainty not stated in the reference; a $\pm 5$~per cent
uncertainty has been assumed.

$^c$The original measurement has been rescaled so that it is
consistent with the current ATCA flux density scale.

$^d$General-width fit; the fixed-width fit has a similar flux density
($1100 \pm 60$ mJy).
\end{minipage}
\end{table*}

\end{document}